\documentclass{aastex}
\usepackage{lscape}
\usepackage{apjfonts}
\usepackage{graphicx}
\usepackage{natbib}
\usepackage{ulem}
\usepackage{multirow}

\newcommand{\thch}{$^{13}$C}
\newcommand{\twch}{$^{12}$C}
\newcommand\teff{T$_{\rm eff}$}
\newcommand\logg{log $g$}

\newcommand{\twth}{$^{12}$C/$^{13}$C}

\begin{document}
\title{Lithium Abundances in CEMP stars\altaffilmark{1}
}

\author{
Thomas Masseron\altaffilmark{2,3},
Jennifer~A.~Johnson\altaffilmark{2},
Sara Lucatello\altaffilmark{4},
Amanda Karakas\altaffilmark{5},
Bertrand Plez\altaffilmark{6},
Timothy~C.~Beers\altaffilmark{7},
Norbert Christlieb\altaffilmark{8}
}
\altaffiltext{1}
{Based on observations collected at the European Southern Observatory, Paranal, Chile (Proposal Number 076.D-0451A)}
\altaffiltext{2}
{Department of Astronomy, Ohio State University,
140 W.\ 18th Ave., Columbus, OH 43210, USA; 
masseron, jaj@astronomy.ohio-state.edu}
\altaffiltext{3} 
{Institut d'Astronomie et d'Astrophysique, Universit\'e Libre de
Bruxelles, CP 226, Boulevard du Triomphe, B-1050 Bruxelles, Belgium }
\altaffiltext{4}
{INAF, Osservatorio Astronomico di Padova, vicolo dell'Osservatorio 5, 35122 Padova, Italy ; Excellence Cluster Universe, Technische Universität München, Boltzmannstr. 2, D-85748 Garching, Germany ; Max-Planck-Institut für Astrophysik, D-85741 Garching, Germany}
\altaffiltext{5}
{Research School of Astronomy \& Astrophysics, The Australian National University, Mount Stromlo Observatory, Cotter Road, Weston, ACT 2611, Australia ; Centre for Stellar and Planetary Astrophysics, School of Mathematical Sciences, Monash University, Clayton, VIC 3800, Australia}
\altaffiltext{6}
{LUPM cc072, Université Montpellier II, F-34095 Montpellier cedex 5 }
\altaffiltext{7}
{Department of Physics \& Astronomy and JINA: Joint Institute for Nuclear
Astrophysics, Michigan State University, E. Lansing, MI 48824, USA}
\altaffiltext{8}
{Zentrum f\"ur Astronomie der Universit\"at Heidelberg, Landessternwarte,
K\"onigstuhl 12, 69117, Heidelberg, Germany}

\begin{abstract}

Carbon-enhanced metal-poor (CEMP) stars are believed to show the
chemical imprints of more massive stars (M$\gtrsim 0.8\,M_\odot$) that are now
extinct. In particular, it is expected that the observed abundance of Li should
deviate in these stars from the standard Spite lithium plateau. We study
here a sample of 11 metal-poor stars and a double-lined spectroscopic binary with $\rm -1.8 <[Fe/H]< -3.3$ observed with
VLT/UVES spectrograph. Among these 12 metal-poor stars, there are 8 CEMP stars
for which we measure or constrain the Li abundance. In contrast to previous
arguments, we demonstrate that an appropriate regime of dilution permits the
existence of ``Li-Spite plateau and C-rich'' stars, whereas some of the
"Li-depleted and C-rich'' stars call for an unidentified additional depletion
mechanism that cannot be explained by dilution alone.  We find evidence
that rotation is related to the Li depletion in some CEMP
stars.

Additionally, we report on a newly recognized double-lined spectroscopic binary
star in our sample. For this star, we develop a new technique from which
estimates of stellar parameters and luminosity ratios can be derived based on a
high-resolution spectrum alone, without the need for input from evolutionary
models.

\end{abstract}

\keywords{ Stars: abundances
Stars: AGB and post-AGB
Stars: Population II
Stars: carbon
(Stars:)binaries: spectroscopic
 }

\section{Introduction
\label{sec:intro}}

Observations of the most metal-poor stars in the Galaxy offer the opportunity to
study the earliest star formation episodes, either through the nucleosynthesis
that polluted the stars under study or through the characteristics of the
surviving stars themselves. One of the striking observational facts established through extensive surveys for metal-poor stars (most notably,
the HK survey, \citeauthor{Beers1992} \citeyear{Beers1992}, and the Hamburg/ESO
Survey, \citeauthor{Christlieb2008} \citeyear{Christlieb2008}) is the large
number of C-rich stars, with as much as 25\% of the stars with [Fe/H] $< -2.5$ having
[C/Fe] $>$ +1.0)~ \citep{Lucatello2005IMF,Lucatello2006}. The origin of these
carbon-enhanced metal-poor \citep[CEMP, ][]{Beers2005} stars is important for
understanding conditions in the early Universe, because the competing theories
predict pollution from different nucleosynthetic sources. Some CEMP stars are
enhanced in the ``slow neutron-capture'' (the $s$-process) elements,
whereas other CEMP stars are also enhanced in the ``rapid neutron capture'' (the
$r$-process) elements. The $s$-process elements are produced primarily in
low-mass asymptotic giant branch \citep[AGB; e.g. ][]{Herwig2005} stars, while the
r-process likely occurs in explosive events such as core collapse supernovae
(SN) \citep[see review by][]{Sneden2008}. CEMP stars with a pure $s$-process
signature are classified as CEMP-s; the subset of CEMP-s stars that possess
evidence for the addition of $r$-process elements are referred to as CEMP-rs
(CEMP-r/s in the original nomenclature of Beers \& Christlieb 2005). Thus far,
only a single example of a CEMP-r star (a class that exhibits pure $r$-process
enrichment, with no apparent contribution from the $s$-process) has been identified,
CS~22902-052 \citep{Sneden2003}. A final class, the CEMP-no stars, exhibit no
enhancements in the neutron-capture elements compared to the average non-C-rich
metal-poor stars in the Galaxy \citep{Beers2005}. 

Many CEMP stars exhibit radial velocity variations, indicating that they are
members of multiple systems such as binaries. A statistical analysis of radial
velocity variations \citep{Lucatello2005bin} and agreement between the observed
abundances of neutron-capture elements and the predictions of $s$-process
production in AGB stars \citep[e.g.,][]{Masseron2010} indicate that the CEMP-s
stars are likely to have all been created by mass transfer from an AGB companion
star. Models for the pollution of the CEMP-rs stars are more uncertain, because
the source of some of their elements, such as Eu, is unclear. Several theories
have been proposed that tie the production of the $s$-process and the
$r$-process together, including accretion-induced collapse of a white dwarf
\citep[see review in ][]{Jonsell2006}. Similarly, the CEMP-no abundance patterns have
been suggested to arise from a number of alternative scenarios.
Proposals for their formation include pollution by a failed supernova
\citep{Umeda2003}, winds from the surface of massive, rapidly rotating, mega
metal-poor ([Fe/H] $< -6.0$) stars in the early Universe \citep{Hirschi2007}, or
by an AGB companion that produced little or no s-process elements \citep{Suda2004,
Masseron2010}.

The Li abundances of CEMP stars are interesting to consider; when compared with
the predictions of theories of C-production, the Li abundances of some CEMP
stars exhibit unexpected results. In non-C-rich, metal-poor stars, stars within
a narrow mass range have sufficiently small convective envelopes that they
preserve most of the Li that was present in the gas clouds from which they
formed, although some Li may be lost to diffusion and turbulent mixing
\citep[e.g., ][]{Richard2005,Melendez2010}. These stars define the Spite plateau
\citep{Spite1982}, and are at or near the main-sequence turnoff in a metal-poor population.
As stars leave the main sequence and become subgiants, the Li preserved in their
outer atmospheres is diluted when it is mixed with material from deeper in the
stellar envelope, where the Li has been burned. As the convective envelope
deepens, and Li is carried down to temperatures high enough to burn, the
stellar envelope becomes increasingly Li-poor.

\begin{figure}[h!]
\includegraphics[width=12cm,angle=-90]{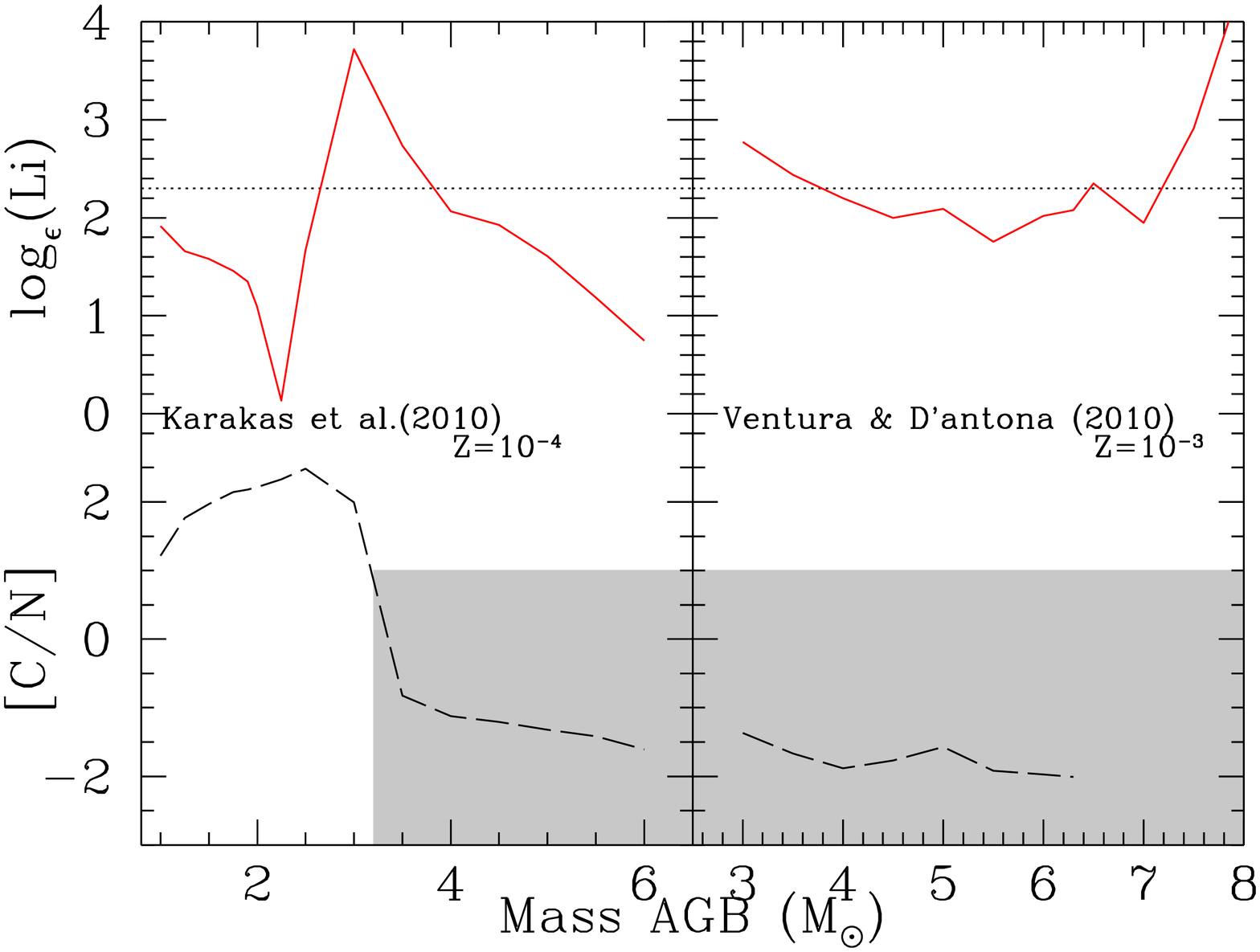}
\caption{(Top panels) Li yields as a function of AGB mass from \citet{Karakas2010} (left panel) 
and from \citet{Ventura2010} (right panel). The dotted line represents the Spite
Li plateau as observed in non-CEMP stars. (Bottom panels) The [C/N] ratio as a
function of AGB mass for the same models. The shaded area indicates the region
where hot bottom burning occurs, which leads to the production of N-rich, rather
than C-rich stars, and therefore will not result in the formation of CEMP
stars.}
\label{fig:Livsmass_Kar_Ventura}
\end{figure}

Lithium may be produced during the AGB phase by the Cameron-Fowler
mechanism, where $^{7}$Be is created at the bottom of the convective envelope
and captures an electron. The resulting Li is carried to the cooler upper layers
of the star before the $^7$Li is destroyed by proton capture
\citep{Sackmann1992}. Gas containing this Li can be released into the
interstellar medium when the AGB star loses mass. However, the predicted production
strongly depends on model assumptions, in particular on the adopted convection
parametrization and mass-loss recipe. Studies of the red giant branch show
that extra-mixing mechanisms can possibly lead to Li production
\citep[][]{Denissenkov2004}, and such extra-mixing mechanisms undoubtedly exist
during the AGB phase as well (e.g., cool bottom processing,
\citealt{Nollett2003,Dominguez2004}, or thermohaline mixing,
\citealt{Stancliffe2010,Charbonnel2010}). Overall, existing yields for Li from
AGB models show a strong mass dependence (see
Fig.~\ref{fig:Livsmass_Kar_Ventura}), and in general, the Li/H ratio in AGB
yields is smaller than the Spite plateau \citep[e.g., ][]{Karakas2010,
Ventura2010}. Models of AGB stars with a range of masses show that Li is
produced at abundance levels greater than the Spite plateau over a limited mass range
(Fig.~\ref{fig:Livsmass_Kar_Ventura}). For the Karakas models, this is M$\sim$
3\,M$_\odot$, while \citet{Ventura2010} find strong Li production only in the
most massive (M$> 7\,M_\odot$) AGB stars. 
The Ventura~\&~D'Antona yields and the Karakas yields are for two different
metallicities, which might explain part of the discrepancy. 
A simple prediction would be that if the CEMP stars were
polluted by AGB stars, the Li abundances of the CEMP stars should reflect the
fact that Li yields from AGB stars have been dumped on them. Another expectation
would be that the Li and C+N abundances in such stars could be used to constrain
nucleosynthesis in low-mass, low-metallicity AGB stars. However, although models accounting for binary interactions are being developped \citep[e.g.][]{Siess2011}, it must be stressed that the impact of the presence of a companion on AGB nucleosynthesis and in particular Li yields is currently not well-constrained.
In the pollution coming from rotating massive stars, some Li depletion is expected (as
well as carbon enhancement). Unless the synthesized material is mixed with large amount of the surrounding ISM, 
log$\epsilon$(Li) will be less than 2.0
\citep{Meynet2010}.

The strongest Li line is the doublet at 6707\AA, a much longer wavelength than
most of the atomic lines that are of interest in CEMP stars. Therefore, in the
case of many observations carried out to date, the Li line was not covered in
the observed part of the spectrum (or the observations were of insufficient
quality to measure this weak feature), thus Li measurements for CEMP stars are
somewhat limited. \citet{Norris1997Cstars} discussed the Li measurements of
\citet{Thorburn1992} and \citet{Thorburn1994} for the CEMP stars LP~625-44,
LP~706-7, CS~22898-027, and CS~22958-042. Both LP~625-44 and CS~22958-042 have Li
abundances lower than non-CEMP stars at a similar evolutionary state, but
LP~706-7 and CS~22898-027, both near the turnoff, have Li abundances very close
to the Spite plateau. The Li abundance in LP~706-7, log$\epsilon$(Li)=2.25, is
slightly below the plateau values, taken by \citet{Thorburn1994} to be
log$\epsilon$(Li)=2.32, while CS~22898-027 is Li-rich relative to the plateau,
with log$\epsilon$(Li) =2.52. Observations of CH stars, which are the result of
AGB mass transfer in moderately metal-poor ([Fe/H] $\sim-1$) binary systems
\citep{McClure1984}, show that they are Li depleted \citep{Smith1993}.
Therefore, \citet{Norris1997Cstars} argued that a Li abundance near the plateau
was evidence against the binary hypothesis. Other CEMP stars with Li abundances
near the Spite plateau are SDSS~J1036+1212 (log$\epsilon$(Li)=2.21;
\citeauthor{Behara2010} \citeyear{Behara2010}) and CS~31080-095
(log$\epsilon$(Li) =1.73) and CS~29528-041 (log$\epsilon$(Li) =1.71), both from
\citet{Sivarani2006}. 

One extremely interesting additional example of CEMP stars with Li
abundance measurements near the Spite plateau is the double-lined spectroscopic binary
CS~22964-161 \citep{Thompson2008}. The working hypothesis offered by these
authors to account for their observations is that the AGB star was a member of a
triple system and polluted the other two stars. They argued that the Li
abundance (log$\epsilon$(Li)=2.09) had to be the result of Li in the mass
transferred from the AGB star, because otherwise the pollution by Li-free
material would lower the abundance to below the plateau value
\citep[log$\epsilon$(Li)=2.10, ][]{Bonifacio2007}. Based on fits of
predicted AGB yields to the observed abundances of the neutron-capture elements, the
authors
argued that the polluting AGB star had M$< 2$M$_\odot$. Because this mass
is not expected to produce Li by the Cameron-Fowler mechanism in standard
models, they suggested
two possible Li production mechanisms that could work in low-mass AGB stars. 
In a H-flash episode, the temperature in the He intershell
region can reach high enough temperatures to produce Li \citep{Iwamoto2004}. Cool bottom processing might
also be responsible for increased Li production in low-mass stars \citep{Dominguez2004}. 
Examples of Li-depleted CEMP turnoff stars also continued to
be found, for example CS~22958-042 \citep{Sivarani2006}, and the most iron-poor
([Fe/H]$=-5.4$) and highest [C/Fe] star ([C/Fe]=+4) HE~1327-2326, with Li
abundance log$\epsilon$(Li) $<$ 0.62 \citep{Frebel2008}, a significant upper
limit well below the Spite plateau. We finally note that there is one remarkable
metal-poor giant, the CEMP-rs star HKII~17435-00532 with a Spite
Li plateau abundance analyzed by \citet{Roederer2008}. Because Li must
be depleted after the first dredge-up, they claim that in-situ Li production is
required in order to explain such a high observed abundance. 

Observations of Li abundances in CEMP stars have therefore established that
there exists a spread in Li values, even among the turnoff stars. It is an open
question if this spread can be reconciled with the binary AGB mass
transfer scenario, or if the carbon in these stars comes from a different
source that does not require the presence of an AGB companion. 

An additional complication for understanding Li abundances in C-rich stars are
processes that occur in the CEMP star itself that could change the surface
abundances of Li, independently of stellar evolution. \citet{Stancliffe2007}
showed that, because CEMP stars have accreted material that has been enhanced in
AGB nucleosynthesis products, an additional mixing process below the conventional stellar
atmosphere of main-sequence stars should occur, the so-called thermohaline
mixing. Thermohaline mixing could affect both the surface Li and C 
abundances if the mixing reaches sufficiently deep into the star. Rotation is
another efficient way to alter the surface abundance of Li, where 
additional meridional circulation and shear turbulence could destroy Li
\citep[see, e.g., ][ and references therein]{Talon2005}.

This paper is organized as follows. In Section~\ref{sec:data} we briefly review the
abundance analysis techniques for the single-lined spectra in our sample. One
of our stars is a double-lined spectroscopic binary, and we devote
Section~\ref{sec:technique_CS22949-008} to presenting a new technique for the
analysis of such systems. This technique avoids the use of isochrones in
measuring the properties of the primaries and secondaries from a single spectrum.
In Section~\ref{sec:results}, we present the Li abundances and carbon isotope
ratios for our program stars. We also calculate the impact on the observed Li
abundance if material from the AGB star is mixed with material in a CEMP star 
and compare with observations from our sample and the literature. Section 4
briefly summarizes our conclusions.

\section{Observations and Data Analysis
\label{sec:data}}

We observed 13 stars with VLT-UVES \citep{Dekker2000}, under ESO program
076.D-0451A, in service mode during October 2005 and March 2006. 
The sample was chosen from the known CEMP stars in the literature available 
for observation from the southern hemisphere. Our goal was to measure
the Li abundance and the \twth{} ratio, so we used the 390+580nm setting, which
resulted in wavelength coverage of 3215\AA~ to 6815\AA. We tried to include as
many warm CEMP stars as possible, as well as a range of CEMP classes, but because
of the limited number of available CEMP targets, some stars cooler than ideal
were observed. The spectra are high resolution (R$\sim$ 45,000) and have a high
signal-to-noise ratio (S/N$\sim$100 per pixel). We re-reduced the UVES pipeline
data, following the procedures of \citet{MasseronPhD}, to obtain the best possible
signal-to-noise ratio in the final spectra.

\subsection{Model Atmospheres and Linelists}

Our analysis was done using the TurboSpectrum code by \cite{Alvarez1998}.
Temperatures and gravities were derived using standard spectroscopic techniques:
abundances independent of excitation potential for Fe~I lines and ionization
equilibrium between Fe~I and Fe~II lines, respectively. Effective temperatures
were also inferred by fitting the wings of the H$\alpha$ Balmer lines.
Microturbulence velocities were determined by requiring no trend for the
abundances of Fe~I lines versus their equivalent widths. The log $gf$ values for Fe
are identical to those of \citet{Hill2002}. Because CNO abundances may have a
strong impact on the structure of the stellar atmospheres
\citep{MasseronPhD}, 1D MARCS models
\citep{Gustafsson2008} have been built specifically for each star, taking into
account the peculiar abundances. The final stellar parameters are listed in
Table~\ref{tab:parameters}. 

The abundances of Li, C, N ,O, Na, Mg, Ba, Eu, and the C isotope ratios were
measured in these stars using synthesis comparisons for all elements except Fe,
because CEMP-star spectra are usually blended with molecular features. When
observable, O was first determined via the O forbidden line at 630nm~
($\log gf =-9.750 $), in order to ensure proper chemical equilibrium of the C-based
molecules. If O was not observable, or was strongly blended by a telluric line,
a value of  [O/Fe] = +0.4 was assumed. The C and N abundances, and
the $^{12}C/^{13}C$ ratios were then measured, using unsaturated CH, CN, and NH
lines (Plez, private communication). Simultaneously, Na abundances
were measured using the NaI lines at 3302\AA~ and 5682-5688\AA~, and
Mg abundances using the MgI 4057-4167-5528-5711\AA~ lines. Atmosphere models and abundance
determination have been iterated until convergence, because their deviation from
standard composition implies changes in the structure of the stellar atmosphere.
The abundances of Ba and Eu were then obtained, including hyperfine structure
and isotope shifts; the atomic data have been taken from 
\citet{Masseron2006} and \citet{Lawler2001Eu}, respectively.
Finally, we measured Li \citep[adopting the linelist of ][]{Hobbs1999}, assuming that the $^6Li/^7Li$ ratio is the solar value
(0.08). We adopt the solar photospheric abundances of \citet{Asplund2005sol} to
calculate the abundance ratios relative to the Sun. Random errors are presented
in parentheses in the tables, and may be attributed to uncertainties in the
continuum placement, to the noise of the observations, and to errors in the
$\log gf$ values of the lines.

\begin{deluxetable}{lrrrrrrrr}

\tablewidth{0pt}
\tablecaption{Stellar Parameters and Abundances\label{tab:parameters}}
\tablehead{\colhead{Star} & \colhead{\teff} & \colhead{\logg} & \colhead{[Fe/H]} &\colhead{$\xi$} & \colhead{log$\epsilon$(O)} & \colhead{log$\epsilon$(Li)} &\colhead{log$\epsilon$(C)}& \colhead{$^{12}$C/$^{13}$C}\\
\colhead{} & \colhead{(K)} & \colhead{} &\colhead{} & \colhead{(km/s)}  & \colhead{} & \colhead{} &\colhead{}& \colhead{} }
\startdata
\multicolumn{7}{c}{CEMP Stars}\\
CS 22183-015  &    5450  &   3.0 &$-$2.87 (0.15)&  1.5&    <7.5          & <0.7         & 7.9  (0.05) &  7    (1)    \\
CS 22887-048  &    6500  &   3.7 &$-$1.80 (0.16)&  1.8&  <8.0            & <1.4         & 8.1  (0.05) &  3    (0.5)  \\
CS 22898-027  &    6000  &   3.5 &$-$2.49 (0.17)&  1.4&    <7.6          & 2.18  (0.1)  & 8.0  (0.05) &  12   (2)       \\
CS 22947-187  &    5200  &   1.5 &$-$2.56 (0.09)&  1.9&   6.8    (0.2)   & <0.5         & 6.95 (0.05) &  3    (0.5)  \\
CS 22949-008a &    6300  &   4.0 &$-$2.14 (0.04)&  1.5&    ...           & <1.0         & 7.85 (0.05) &  >30         \\ 
CS 22949-008b &    5300  &   4.7 &$-$2.14 (0.04)&  1.5&    ...           & <1.0         & 7.85 (0.05) &  >30         \\
CS 29512-073  &    5600  &   3.4 &$-$2.09 (0.14)&  1.3&   <7.6           & 1.93  (0.1)  & 7.55 (0.05) &  >60        \\
CS 30322-023  &    4100  &$-$0.3 &$-$3.33 (0.19)&  2.2&    5.9   (0.1)   &<$-$0.3         & 5.6  (0.05) &  4    (0.5)  \\
HD 198269     &    4500  &   1.0 &$-$1.84 (0.25)&  1.8&   ...            & <0.2         & 7.45 (0.05) &  5    (1)  \\
\multicolumn{7}{c}{Metal-Poor Stars}\\
CS 22948-104  &    5000  &   2.3 &$-$2.76 (0.11)&  1.6&   <6.7           & 0.92  (0.1)  & 6.05 (0.05) &  >50         \\
CS 29493-090  &    4700  &   0.9 &$-$3.25 (0.15)&  1.6&   6.3    (0.2)   &<$-$0.2         & 5.95 (0.05) &  6    (1)  \\
CS 29517-025  &    5300  &   1.2 &$-$2.57 (0.09)&  2.0&  <6.9            &<0.5          & 6.0  (0.05) &  ...        \\
CS 30312-100  &    5000  &   2.0 &$-$2.70 (0.11)&  1.5&    6.9   (0.2)   & 0.85  (0.1)  & 6.08 (0.05) &  >50         \\
\enddata
\end{deluxetable}

\subsection{Spectroscopic Analysis of the Double-Lined Binary}\label{sec:technique_CS22949-008}

Among our sample, one star (CS~22949-008) is a double-lined spectroscopic binary
(or SB2). Four SB2s composed of metal-poor stars ([Fe/H]$< -1.5$) have previously
received extensive discussions in the literature: CS~22873-139, first analyzed
by \citet{Preston1994bin} and later re-observed by \citet{Spite2000};
CS~29527-015, studied by \citet{Thorburn1994} and \citet{Norris1997a}, with the
Li detection by \citet{Spite2000}; CS~22876-032, studied by \citet{Norris2000}
and \citet{GonzalezHernandez2008}; and CS~22964-161, analyzed by
\citet{Thompson2008}. As mentioned in the Introduction, this latter system
exhibits a large C enhancement, similar to that of our system.

A proper abundance analysis of such a system requires deriving the four
parameters (T$\rm_{eff}$, $\log g$, metallicity, microturbulence, as well as the
convolution profile for proper synthesis) of the two components of the binary
based on information extracted from a single spectrum. To successfully separate
the two stars, the luminosity ratio of the two stars must also be determined.
Standard methods \citep[e.g., ][]{Thompson2008} usually make use of isochrones
to determine this ratio. However, thanks to the large wavelength coverage of our
data, we can derive the stellar parameters for each component, as well as their
luminosity ratio, using only a single high-resolution spectrum, as described in this
section.

\subsubsection{Principles}






By taking into account the veiling flux ($F_B R_B^2$) of star B, the
observed equivalent width $EW^{obs}$ for star A can be expressed such that
$EW^{obs}_A \times (F_A R_A^2+F_B R_B^2) = EW_A \times F_A R_A^2$, where
$F_{\lambda,i}$ is the continuum flux at the wavelength $\lambda$ of star $i$,
 $EW_A$ is the computed equivalent width for star A, and $R_i$ is the
radius of star $i$. This formula can be then rearranged the following way:
\begin{eqnarray} \label{eqn:R_ratio}
\Big(\frac{EW_A}{EW^{obs}_A} - 1 \Big) \times \frac{F_{\lambda,A}}{F_{\lambda,B}}  & = & \Big(\frac{R_B}{R_A}\Big)^2  
\end{eqnarray}
(and reciprocally for component B).

While $EW^{obs}_i$ can be directly measured on the spectrum (if the radial
velocity shift between the two components is sufficiently large), $F_{\lambda,i}$ and
$EW_i$ can be computed from synthetic spectra for any effective temperature,
gravity, metallicity, and microturbulence. Note that using equivalent widths
offers the advantage of being independent of the instrumental, macroturbulent, or
rotation profile. These profiles can be derived by synthetic comparison once all
the other parameters have been set. 

Moreover, Eqn.~\ref{eqn:R_ratio} shows that the appropriate combination of these
factors should be constant, as $(\frac{R_B}{R_A})^2$ represents the radius ratio
of the stars. This property can be used to derive the stellar parameters of the
individual components of an SB2. In particular, $(\frac{R_B}{R_A})^2$ must be
independent of the excitation potential of the spectral lines. As is the case in the
analysis of single-star spectra, the dependence of $EW_A$ on excitation
potential is mostly sensitive to changes in effective temperature. Thus, by
requiring no trend of $\big(\frac{R_B}{R_A}\big)^2$ on excitation potential, we
can constrain the effective temperature of each star.

\begin{figure}[h!]
\includegraphics[width=10cm,angle=-90]{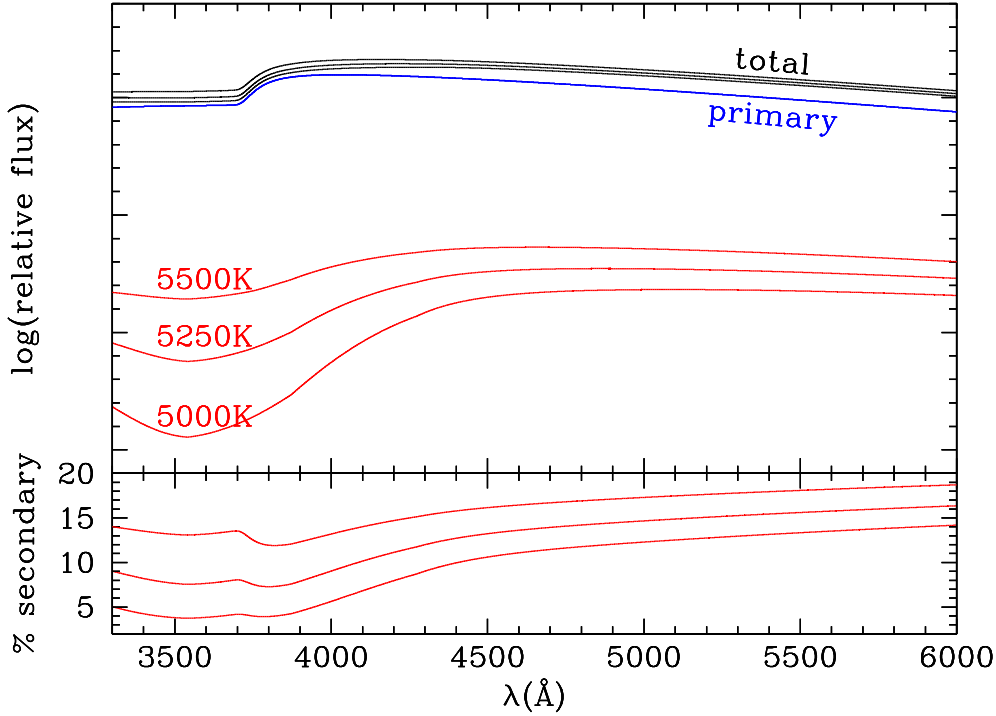}
\caption{Example of the variation of the continuum flux distribution with wavelength for each component of an SB2, for
three temperatures of the secondary, with the radius ratio and the temperature of
the primary fixed to the case of the SB2 CS~22949-008. The discrepancy between
the red and the blue portion of the contribution of the secondary flux to the total
flux (lower panel) is more pronounced for the coolest temperatures.}
\label{fig:flux_contrib}
\end{figure}

In addition, the term $F_{\lambda,A}/F_{\lambda,B}$ in Eqn.~\ref{eqn:R_ratio} is
dependent on wavelength. Fig.~\ref{fig:flux_contrib} illustrates how the
contribution of the flux can vary over the spectrum, depending on its effective
temperature. Hence, effective temperatures can be constrained by requiring
no dependence of Eqn.~\ref{eqn:R_ratio} with wavelength instead of excitation
potential. Because of the dependence of Eqn.~\ref{eqn:R_ratio} on excitation
potential and wavelength and also on the continuum flux of the other component
(i.e., its effective temperature), the temperature determination process must be
iterated until convergence.

\begin{figure}[h!]
\includegraphics[width=10cm,angle=-90]{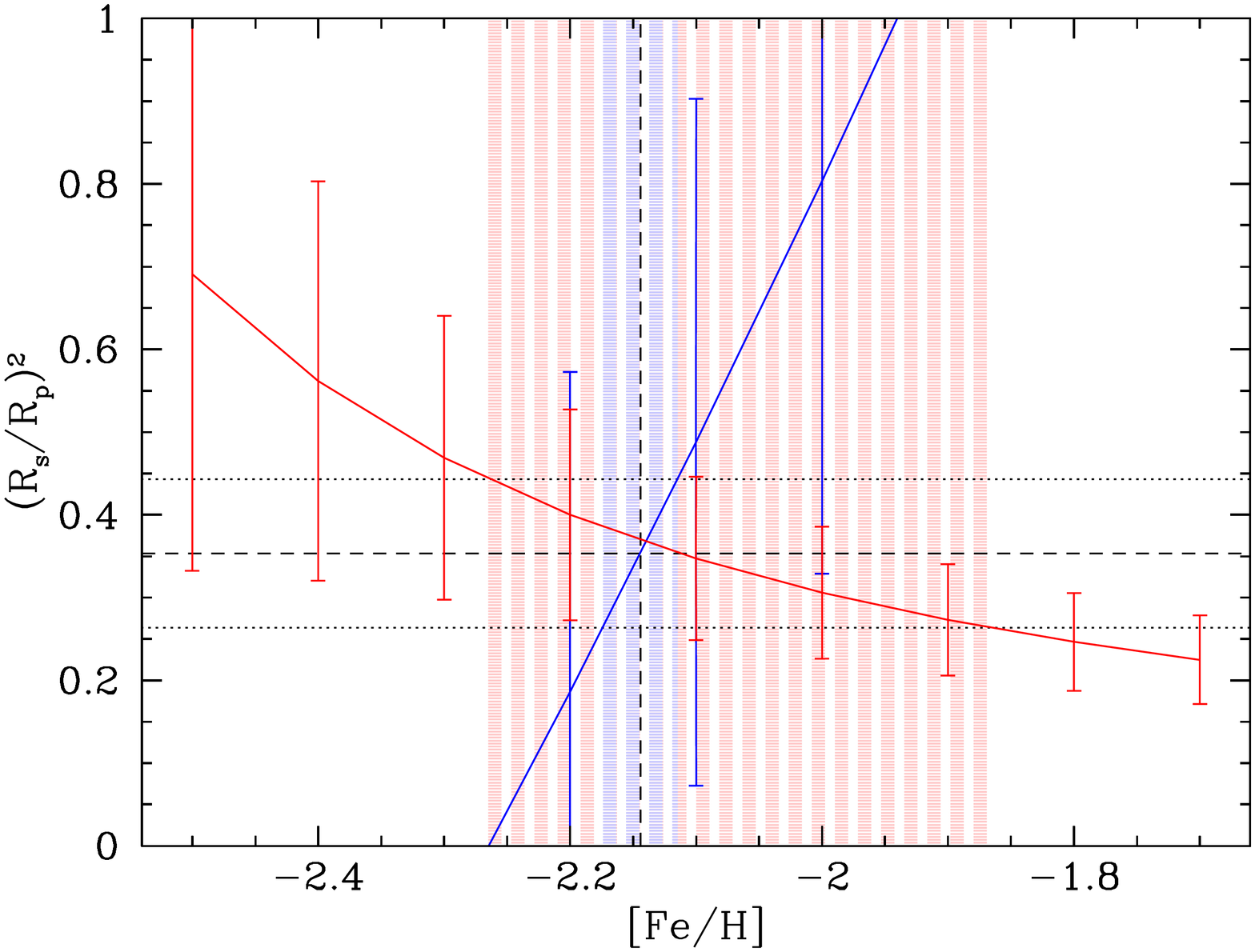}
\caption{Variation of the radius ratio with metallicity, using Eqn.~\ref{eqn:R_ratio} in the
case of the SB2 CS~22949-008, with either the primary's (blue line) or the
secondary's (red line) EW$^{obs}$. The uncertainty on the radius ratio (shown
with dotted lines) is due to the observed scatter in the equivalent widths. The shaded
areas indicate the subsequent error on the derived metallicity. There is a
dependency of the radius ratio on the metallicity.}
\label{fig:RvsFe}
\end{figure}

However, there is no constraint in Eqn.~\ref{eqn:R_ratio} on the abundance. 
Consequently, the derived radius ratio from this equation is dependent on the
metallicity, as illustrated in Fig.~\ref{fig:RvsFe}. Assuming that the
metallicity is identical for both components, the radius ratio and the metallicity
of the system can then be immediately derived simultaneously 
by choosing the metallicity that gives the same radius ratio for both 
the primary and secondary EWs. Without this assumption,
Eqn.~\ref{eqn:R_ratio} is completely degenerate between the ratio of the
metallicities of each component and the radius ratio. In this case, additional
constraints on the radius ratio may be set from their observed orbital motion, or
constraints on metallicity may be derived from another element.

Once the effective temperatures of the secondary and the primary, as well as the
radius ratio, have been fixed, the veiling factor $f_{\lambda,i} =
\frac{EW_i}{EW^{obs}_i}$ can be computed for any set of lines. Hence, the
respective EW$_i$ of individual stars can be recovered and standard techniques
applied. Thus, gravity can be determined from the ionization equilibrium, as is
usually done. Similarly, microturbulence velocity can be set by comparing the
abundances of strong and weak lines. Note that Eqn.~\ref{eqn:R_ratio} is also
dependent on the microturbulence, thanks to $EW_A$, and the stellar parameter determination for
the components has to be iterated until convergence.

In principle, this method should be applicable to any SB2 for which a spectrum
with high resolution and a large wavelength range is available, with the minimum
condition that the radial velocity shift is sufficiently large to permit the
measurement of the equivalent widths of each component separately. In fact, this
technique could even be extended to any SB2 spectrum, once constraints are known
(or adopted) on the metallicity ratio or radius ratio of the components.

\subsubsection{Application to CS~22949-008}\label{sec:rem_features}

We have applied the above technique to our double-lined spectroscopic binary
CS~22949-008, using a set of measured equivalent widths for the Fe~I lines. The derived
parameters are listed in Table~\ref{tab:CS22949-008_parameters}, with errors
derived from the line-to-line scatter (one standard deviation). It may be
counter-intuitive that the error on the effective temperature of the secondary
is better constrained than that of the primary. However, as illustrated in
Fig.~\ref{fig:flux_contrib}, a change of $250$~K in the temperature of the
secondary leads to a change of 5\% in its flux contribution to the total flux.
Since the SNR of our spectrum is typically 100, it follows that the precision of
the flux is known as well as 1\%. Hence, a precision of $\rm \approx 50$~K for the
temperature of the secondary can be achieved, while the uncertainty of the
temperature of the primary is similar to literature values,
because it is essentially dominated by the uncertainty on the spectral line data.

\begin{deluxetable}{l|c|c}
\tablewidth{0pt}
\tablecaption{CS~22949-008 parameters \label{tab:CS22949-008_parameters}}
\tablehead{              & \colhead{Primary}             & \colhead{Secondary}  }
\startdata
T$\rm_{eff}$(K) &$6190^{+130}_{-150}$ & $5250^{+50}_{-50}$\\
$\log$g         &  $4.0 \pm 0.2$      & $ 4.7 \pm 0.3$ \\
$\xi_t$ (km/s)  &1.5 $\pm$ 0.3        & 1.3 $\pm$ 0.5\\
$[Fe/H]$                   & \multicolumn{2}{c}{$-2.14 \pm 0.04$}  \\ 
$\frac{R_{sec}}{R_{prim}}$ & \multicolumn{2}{c}{$0.59 \pm 0.07$}    \\
$\frac{L_{sec}}{L_{prim}}$ & \multicolumn{2}{c}{$0.18 \pm 0.04$ }           \\
\enddata
\end{deluxetable}

Table~\ref{tab:CS22949-008_errors} lists the systematic errors on the derived
parameters of the SB2 due to uncertainties in the atmospheric parameters of each
component. It is noticeable that changes in the stellar parameters of the
primary do not impact on the determination of the parameters of the secondary.
Indeed, the temperature of the secondary is strongly dependent on the
wavelength, whereas a change in temperature of the primary acts as an almost
constant change in the luminosity flux over wavelength, thus it has little
impact on the temperature determination. Overall, the technique we employ
allows us to estimate the stellar parameters of CS~22949-008 fairly
precisely. However, we emphasize that there is a relatively large contrast in
effective temperature between the two components, such that
Eqn.~\ref{eqn:R_ratio} is mostly sensitive to excitation potential in the case of
the primary, whereas it is mostly sensitive to wavelength variation for the
secondary. Hence, the derivation of stellar parameters for the two components is
quite independent of one another. Indeed, the degeneracy that can be
lifted in the analysis of our system, with a rather large contrast in
temperature, is likely more robust than for systems with similar
temperatures of the components.

\begin{deluxetable}{l|c|c|c|c}
\tablewidth{0pt}
\tablecaption{Error Budget for CS~22949-008 \label{tab:CS22949-008_errors}}
\tablehead{                     &\multicolumn{2}{c|}{Primary}    & \multicolumn{2}{c}{Secondary}      \\
                   &  \colhead{T$\rm _{eff}+150K$}  &  $\xi_t$+0.5   & \colhead{T$\rm _{eff}+150K$} &  \colhead{$\xi_t$+0.5}   }
\startdata
$\Delta$T$\rm_{eff\_prim}$         & {\it +150}   &     {\it +0.5}    &     +15         &      0          \\
$\Delta$T$\rm_{eff\_sec}$          & +16          &    0              &   {\it +150}    &    {\it +0.5}   \\
$\Delta$$[Fe/H]$                   &   +0.11      &  -0.06            &  +0.02          &     -0.01       \\
$\Delta$$\frac{R_{sec}}{R_{prim}}$ &  -0.053      &  +0.015           & -0.007          &    -0.014      \\
$\Delta$$\frac{L_{sec}}{L_{prim}}$ &  +0.02       &  -0.01            & -0.02           &    +0.01        \\
\enddata
\end{deluxetable}

\begin{figure}[h!]
\includegraphics[width=10cm,angle=-90]{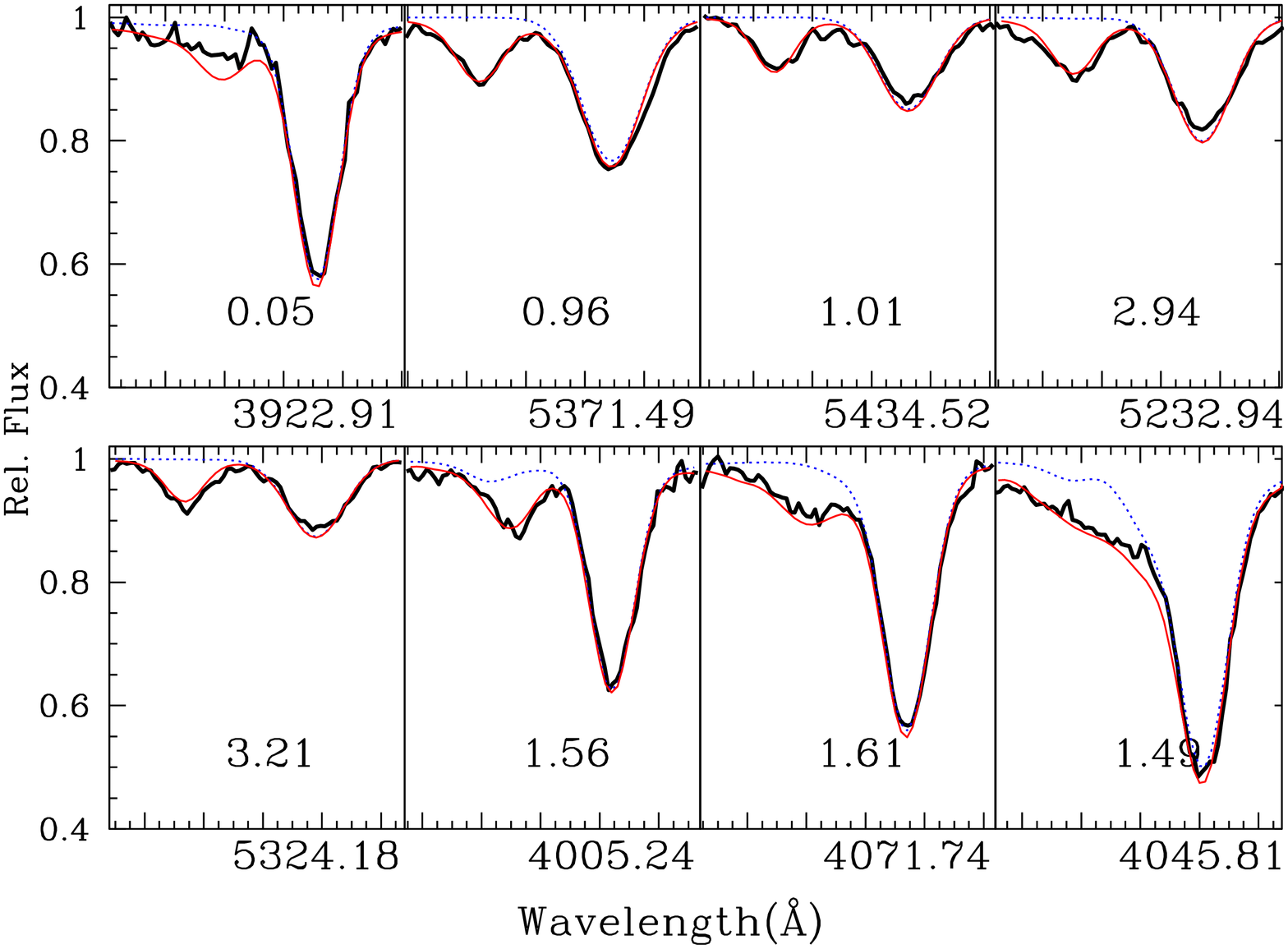}
\caption{Fit to the Fe~I lines of the combined spectrum of the SB2 CS~22949-008. 
The thick black line shows the observed spectrum, while the thin red line is the
synthetic fit adopting the previously determined parameters, and the blue dotted
line represents the contribution to the combined spectrum of the primary star.
The labels show the respective excitation potential of each line.}
\label{fig:FeI_CS22949}
\end{figure}

Fig.~\ref{fig:FeI_CS22949} presents examples of the Fe~I line fits. 
The radial velocity shift between the two stars is 25\,km/s.  From this fit,
convolution profiles have been set to 12\,km/s for the primary star
and 6\,km/s  for the secondary star.

\begin{figure}[h!]
\includegraphics[width=12cm,angle=-90]{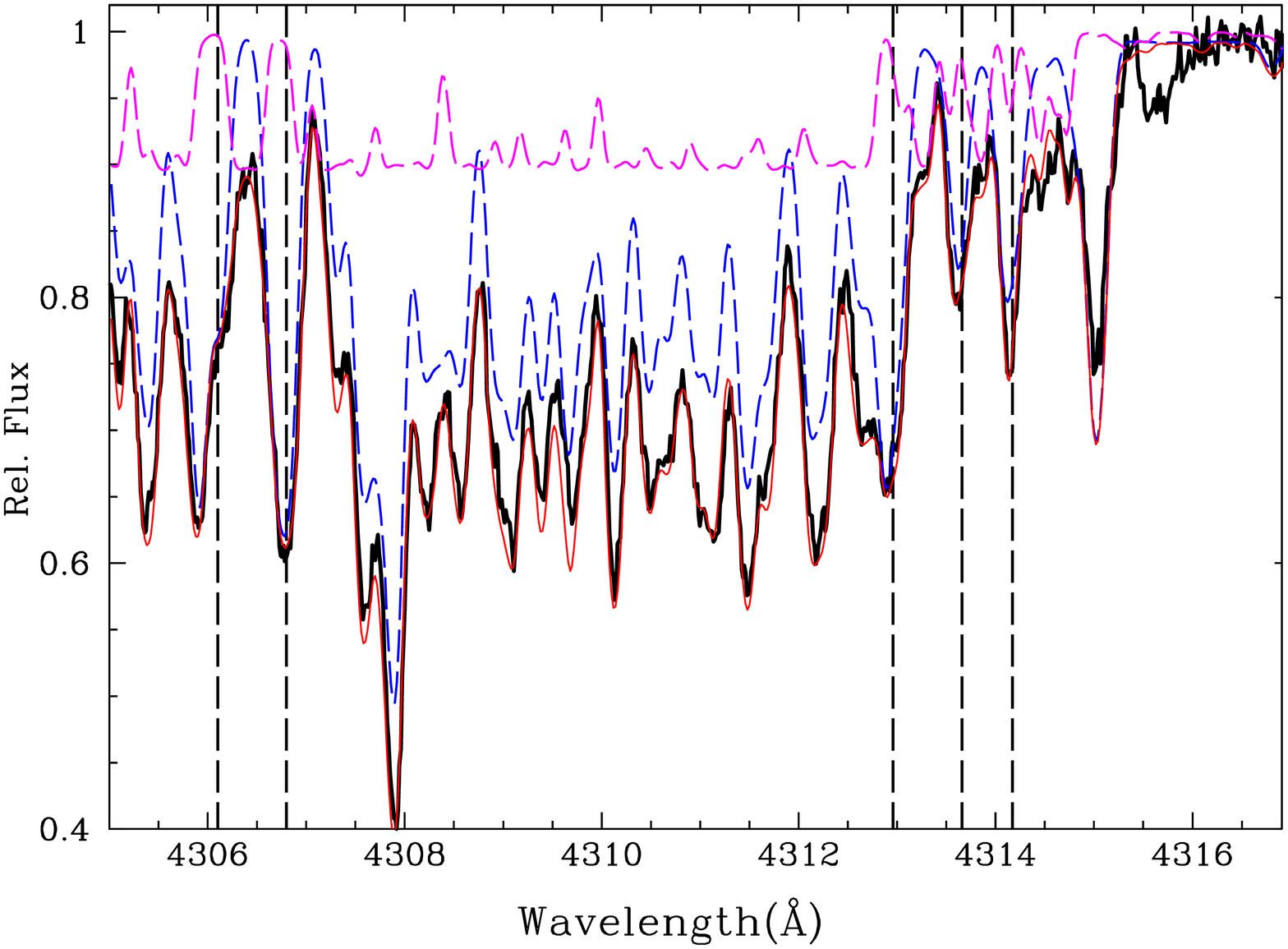}
\caption{The fit of the CH lines with $\log_\epsilon(C)=7.85$. The black thick line is the
observed spectrum of CS~22949-008, while dotted lines mark the respective contribution of the primary and the secondary. 
Vertical dashed lines indicate where the CH lines are more trusted for the determination of the C abundance for the primary.}
\label{fig:CH_CS22949}
\end{figure}
\begin{figure}[h!]
\includegraphics[width=5cm,angle=-90]{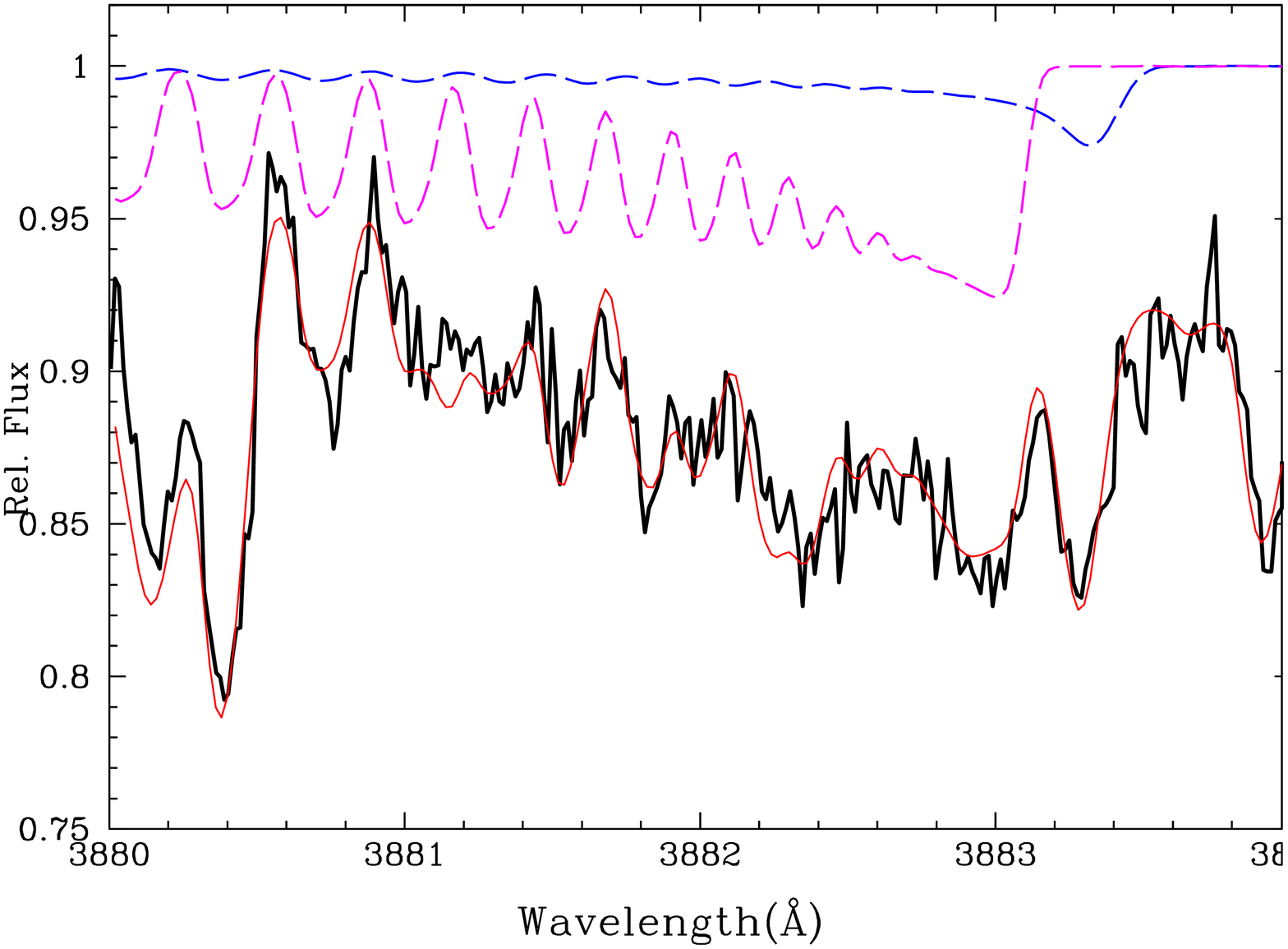}
\includegraphics[width=5cm,angle=-90]{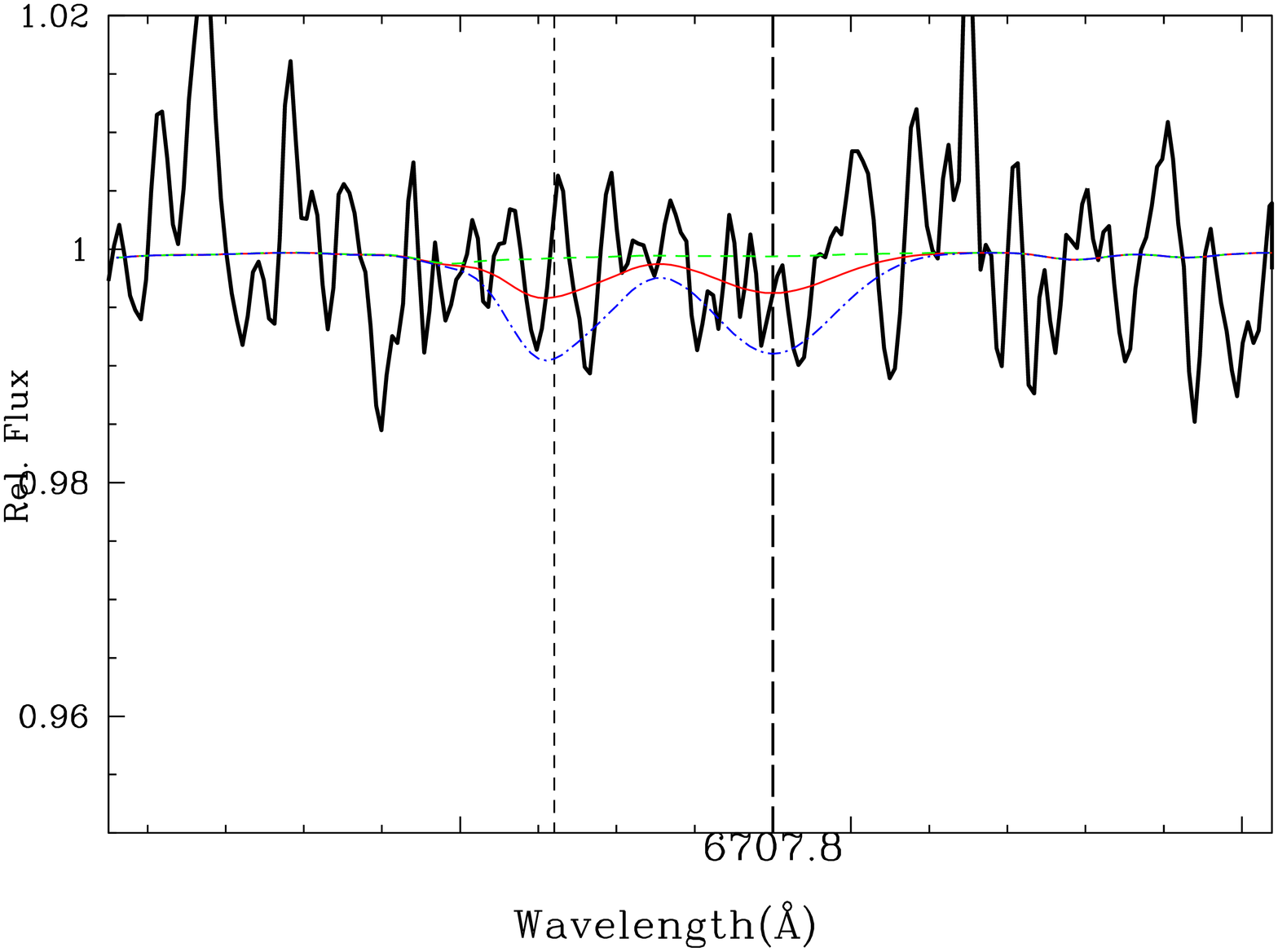}
\caption{CN band fit and Li line fit. Left panel: the dotted blue and magenta lines illustrate
the respective contribution of the primary and the secondary CN bands. The solid red line represents the total synthetic fit with $\log_\epsilon(N)=6.1$.
 Right panel:  $\log_\epsilon(Li)=0.0,1.0,1.4$
(green dashed line, solid red line, and blue dash-dotted line, respectively).
The vertical long-dashed line shows the position of the Li line of the primary
and the short-dashed line the position of the Li for the secondary. }
\label{fig:CN+Li_CS22949}
\end{figure}

Because of the low temperature of the secondary, its spectral lines are often
saturated, but not when their depth is above $\sim$90\% of
the continuum of the combined spectrum. For example, Fig.~\ref{fig:CH_CS22949}
shows that the lines of the CH G-band are due exclusively to the primary, while the
secondary acts only as a continuum offset. Hence, for our abundance analysis, a
careful selection of lines has been performed, so that the chosen lines reach
less than $\sim$90\% of the continuum of the combined spectrum (e.g.,
Fig.~\ref{fig:CH_CS22949}). In contrast, because the primary star is warmer, the
CN (as well as C$_2$) bands appear mostly because of the secondary contribution to
the spectrum (Fig.~\ref{fig:CN+Li_CS22949}). The Li line does not appear in
either star's spectra. Therefore, we fix the upper limit to $\log_\epsilon(Li) <
1.0)$ (Fig.~\ref{fig:CN+Li_CS22949}).

\begin{figure}[h!]
\includegraphics[width=10cm,angle=-90]{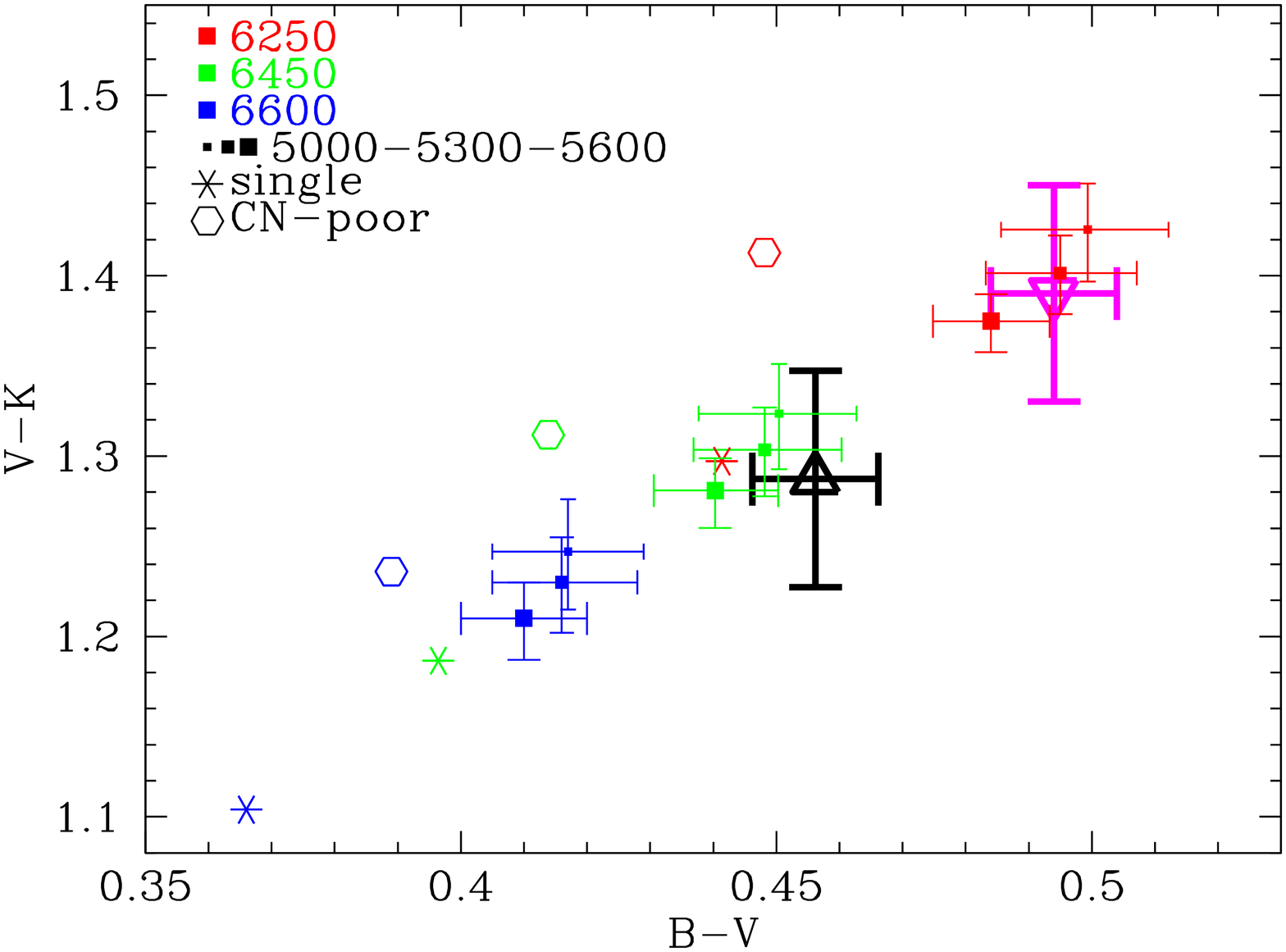}
\caption{V-K versus B-V color of CS~22949-008, both corrected for
reddening (black open up triangle) and uncorrected for reddening (open magenta
down triangle). The squares, hexagons, and stars are synthetic colors
for the binary C-enhanced system, the binary non-C-enhanced system, and
the C-enhanced single system, respectively. There is fairly good agreement
between the uncorrected colors and those predicted using synthetic spectra for
a C-enhanced system with binary stars of temperature 6200~K and 5300~K.}
\label{fig:VKvsBV_CS22949}
\end{figure}

Using the previously determined radius ratio, we were able to compute synthetic
colors for CS~22949-008 (Fig.~\ref{fig:VKvsBV_CS22949}). Note that we had to
compute specific atmosphere models and synthetic colors to properly take into
account the effect of C on the photometry \citep[see ][]{MasseronPhD}. We see in
Fig.~\ref{fig:VKvsBV_CS22949} that, although the secondary's temperature cannot
be derived from this diagram, it contributes to the B-V and V-K colors, and tends
to ``cool down'' the apparent color compared to a single star by 200~K.
Synthetic colors using the stellar parameters determined as described above
agree fairly well with the observed colors, although a better match is obtained
when the colors are not corrected for reddening. Indeed, we note in
\citet{Masseron2010prep} that the match between spectroscopic and photometric
temperatures is generally better with the observed colors.  

\begin{figure}[h!]
\includegraphics[width=10cm,angle=-90]{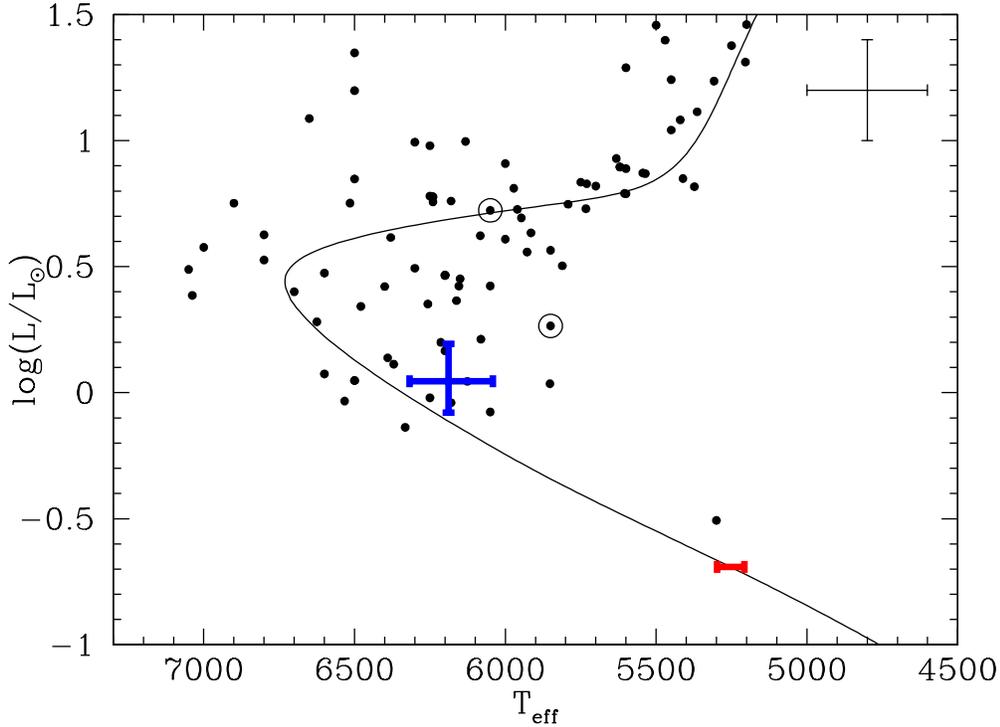}
\caption{HR diagram including the two components of the SB2 
CS~22949-008 (blue cross and red line). Black dots are CEMP stars from the 
literature, for which the luminosity has been calculated from their T$\rm_{eff}$
and $\log g$, assuming $M=0.8M_\odot$ such that $\rm \log L/L_\odot= \log (M/M_\odot) + 4 \log(T_{eff}/T_{eff\odot})-\log (g/g_\odot)$.
The dotted circles represent the other known C-enhanced SB2 star,
CS~22964-161 \citep{Thompson2008}. The luminosity of the secondary has been
set according to the spectroscopic effective temperature combined
with an isochrone for 12\,Gyr and $\mathrm{[Fe/H]}=-2.3$ \citep[][thin solid line]{VandenBerg2006}.
The luminosity of the primary has then been computed
using the radius ratio and the effective temperature
determined spectroscopically in Sect.~\ref{sec:technique_CS22949-008}.}
\label{fig:binaryiso}
\end{figure}

For this system, the luminosity ratio determined via the spectroscopic technique
does not exactly match the one predicted by standard evolutionary models
(Fig.~\ref{fig:binaryiso}). While $\log g$ for the secondary is in agreement
with the theoretical value ($\log g = 4.7$), the $\log g = 4.0\pm0.2$ for the
primary is lower than the value of $\log g =$4.5 expected based on the
isochrones. Although we carefully estimate errors in
Table~\ref{tab:CS22949-008_parameters} and \ref{tab:CS22949-008_errors}, it
should be stressed that effects such as NLTE can significantly impact
estimates of surface gravity. Although we cannot evaluate quantitatively these
effects at present, a reasonable change of $\approx 0.2$~dex in gravity would lead to an
agreement between the observed and theoretical luminosity ratio to within the
uncertainties. We note that \citep{Thompson2008} found a similar
problem in the other known C-rich metal-poor SB2 star, CS~22964-161, for which an
orbital solution exists and a mass ratio could be derived. The resulting
luminosity ratio in this system appeared to also be inconsistent with the
gravity derived from spectroscopy (Fig.~\ref{fig:binaryiso}).    

\newpage

\section{Results and Discussion}\label{sec:results}

\subsection{Abundances}

Tables \ref{tab:parameters} and \ref{tab:Abundances} present the derived
abundances or limits for Li, C, N, O, the $^{12}$C/$^{13}$C ratio, Na, Mg, Ba,
and Eu for the stars in our sample. Parentheses indicate the random errors on
the reported abundances. Table \ref{tab:errors} provides estimates of the errors
of the abundances due to the uncertainties in the adopted stellar parameters for
two typical stars from our sample, one dwarf and one giant. Five of the stars in
our sample have been observed previously. Table \ref{tab:Litcomp} compares our
atmospheric parameters and abundances with these results. The stellar parameters
found for the dwarfs generally agree with the published values within the error
bars. However, there are larger discrepancies for the giants, which may be
attributed to the differences in the adopted stellar atmosphere models, notably
the inclusion of the effect of enhanced C in our study. We have complemented the
Li abundances measured in this paper with Li abundances reported for CEMP stars
in the literature, summarized in Table~\ref{tab:extended}. Note that we use the
published abundances, without application of any modifications for the solar
reference abundances or NLTE corrections. 

\begin{deluxetable}{lrrrrrc}

\tablewidth{0pt}
\tablecaption{Abundances\label{tab:Abundances}}
\tablehead{\colhead{Star} & \colhead{log$\epsilon$(N)} & \colhead{log$\epsilon$(Na)} & \colhead{log$\epsilon$(Mg)} & \colhead{log$\epsilon$(Ba)} & \colhead{log$\epsilon$(Eu)}  &  \colhead{subclass}}
\startdata
\multicolumn{7}{c}{CEMP Stars}\\
CS 22183-015  & 7.05  (0.05)  &  3.70  (0.1)  & 5.25 (0.05) &  1.2  (0.1)  & $-$0.85  (0.1) & rs \\
CS 22887-048  & 7.40  (0.05)  &  4.60  (0.05) & 6.00 (0.00) &  2.1  (0.1)  &  0.1   (0.1) & rs \\
CS 22898-027  & 6.75  (0.05)  &  4.05  (0.05) & 5.45 (0.1)  &  2.2  (0.1)  & $-$0.1   (0.05)& rs \\
CS 22947-187  & 7.00  (0.05)  &  3.80  (0.05) & 5.50 (0.05) &  0.9  (0.1)  & $-$1.4   (0.1) & s \\
CS 22949-008p & 6.10  (0.1)   &  5.00  (0.1)  & 6.0  (0.1)  &  1.4  (0.1)  &$<-$1.0        & s \\ 
CS 22949-008s & 6.10  (0.1)   &  ...          & 6.0  (0.1)  &  1.4  (0.1)  &$<-$0.2        & s \\
CS 29512-073  & 6.30  (0.05)  &  4.10  (0.05) & 5.70 (0.1)  &  1.25 (0.1)  & $-$1.3   (0.1) & s \\
CS 30322-023  & 7.20  (0.05)  &  4.10  (0.05) & 4.94 (0.05) & $-$0.7  (0.1)  & $-$3.5         & low-s \\
HD 198269     & 7.20  (0.1)   &  4.35  (0.05) & 6.15 (0.1)  &  1.6  (0.1)  & $-$0.8   (0.1) & s \\
\multicolumn{7}{c}{Metal-Poor Stars}\\
CS 22948-104  & 5.10  (0.1)   &  3.35  (0.2)  & 5.15 (0.05) & $-$0.85 (0.1) &$-$2.4   (0.1) & r-I \\
CS 29493-090  & 5.90  (0.05)  &  3.75  (0.05) & 5.30 (0.05) & $-$0.7  (0.1) &$<-$3.5        & r-I \\
CS 29517-025  & 6.60  (0.1)   &  3.90  (0.05) & 5.45 (0.05) & $-$0.6  (0.1) &$-$1.5   (0.1) & r-I \\
CS 30312-100  & 5.00  (0.2)   &  3.50  (0.1)  & 5.35 (0.05) & $-$1.30 (0.05)&$-$2.3   (0.1) & r-I \\
\enddata
\end{deluxetable}

Adopting the same definition of CEMP stars as in \citet{Masseron2010}
([C/Fe]$\ge$+0.9), it appears in this sample that five stars of our sample are
not CEMP stars (CS~30312-100, CS~22948-104, CS~29493-090, CS~29517-025,
and CS~30322-023). Although these stars were originally selected as CEMP stars from
the low-resolution survey of \citet{Beers1992}, the present high-resolution
study reveals that they have been misclassified, possibly due to their
relatively cool effective temperature causing uncertainties in the automatic
classification. Therefore, these stars will not be considered as CEMP stars
in the following discussion except CS~30322-023 that we kept as a low-s CEMP star as argued in \citet{Masseron2006,Masseron2010}. However, we note that the cool effective temperature of  CS~30322-023 and its extremely low
gravity suggest that this star has sufficiently deep convection zones to have at
least partially destroyed its surface Li, complicating the discussion of the
origin of its Li.

\begin{deluxetable}{l|ccc|ccc}

\tablewidth{0pt}
\tablecaption{Errors on Abundances \label{tab:errors}}
\tablehead{$\Delta \log_\epsilon(X)$ & & CS~22898-027 & & & CS~30312-100& } 
\startdata  
            &T$\rm_{eff}$ + 150 & $\log g$ + 0.2 & $\xi_t$ + 0.2 & T$\rm_{eff}$+ 150 & $\log g$ + 0.2 & $\xi_t$ + 0.2 \\
Li          &     0.11          &     0.03     &   0.03        &         0.12       &   $-$0.01      &   0.00      \\ 
C           &     0.30          &     0.01     &   0.07        &         0.34       &   $-$0.06      &  $-$0.01      \\ 
N           &     0.30          &     0.05     &   0.06        &         0.30       &   $-$0.08      &   0.02      \\ 
O           &     0.08          &     0.10     &   0.03        &         0.07       &    0.04      &   0.00      \\ 
Na          &     0.06          &     0.02     &   0.02        &         0.07       &    0.00      &   0.01      \\ 
Mg          &     0.07          &     0.05     &   0.02        &         0.09       &   $-$0.01      &  $-$0.02      \\ 
Fe          &     0.14          &     0.03     &   0.00        &         0.17       &   $-$0.01      &  $-$0.05      \\ 
Ba          &     0.10          &     0.08     &   0.01        &         0.09       &    0.06      &  $-$0.03      \\ 
Eu          &     0.10          &     0.10     &   0.03        &         0.09       &    0.09      &   0.00      \\ 
\enddata
\end{deluxetable}

\begin{deluxetable}{lrrrrrr}

\tablewidth{0pt}
\tablecaption{Comparison with Previous Results \label{tab:Litcomp}}
\tablehead{\colhead{Star} & \colhead{\teff} & \colhead{\logg} &
\colhead{[Fe/H]} &\colhead{log$\epsilon$(Li)} & \colhead{[C/Fe]} 
&\colhead{Source} }
\startdata
CS 22183-015  &  5200&2.5  &$-$3.12  & ... & +2.2  & (4)  \\
              &  5733&3.6  &$-$2.37  & ... & +2.42 & (5) \\
              &  5620&3.4  &$-$2.75  & ... & +1.95 & (6)   \\      
              &  5470&2.85 &$-$2.85  & ... & +2.34 & (2)  \\           
              &  5450&3.0  &$-$2.87  & <0.7 & +2.33 & (8)  \\           
CS 22887-048  &  6500&3.35 & $-$1.70 & ... & +1.84 & (2)  \\
              &  6500&3.70 &$-$1.75  & <1.4 & +1.46 & (8)  \\           
CS 22898-027  &  6250&3.7  & $-$2.26 &2.10 & +2.2  & (1)  \\
              &  6240&3.72 & $-$2.30 & ... & +2.34 & (2)  \\
              &  6300&4.0  & $-$2.00 & ... & +1.95 & (3)  \\
              &  6000&3.5 &$-$2.44  & 2.18 & +2.05 & (8)  \\           
CS 30312-100  &  5300&2.8  & $-$2.33 &<1.6 &...   & (1)  \\
              &  5000&2.0  &$-$2.65  & 0.85 & +0.34 & (8)  \\ 
CS 30322-023  &  4300&1.00 & $-$3.25 & ... & +0.56 & (7)  \\
              &  4100&-0.3 &$-$3.28  & <$-$0.3 & +0.49 &  (8)  \\           
\enddata
\tablerefs{(1) \citet{Aoki2002c}, (2) \citet{TsangaridesPhD}, (3) \citet{Preston2001},
(4) \citet{Johnson2002}, (5) \citet{LucatelloPhD}, (6) \citet{Cohen2006}, (7) \citet{Aoki2007}, (8) This work}

\end{deluxetable}

\subsection{Evolutionary Status }

As indicated in the Introduction, whatever Li is present at the surface of a
CEMP star, either original or deposited, will be affected by the evolution of
that star. Therefore, to interpret the observed Li abundances, the evolutionary
status of the CEMP star must be known. However, in the phases where Li is
fully destroyed (notably during the AGB and in the Li-dip), the origin of the
Li (if there was any present originally) cannot be recovered. Thus, stars in these phases will be disregarded
for the purpose of our discussion.
{\small
\begin{deluxetable}{lrrrrrrrrr}
\tabletypesize{\small}
\tablewidth{0pt}
\tablecaption{The Extended Sample\label{tab:extended}}
\tablehead{\colhead{Star} & \colhead{\teff \, \logg} & \colhead{[Fe/H]} &\colhead{[C/Fe]} & \colhead{[Ba/Fe]} &\colhead{$^{12}$C/$^{13}$C}&\colhead{[N/Fe]}&\colhead{log$\epsilon$(Li)} & subclass & \colhead{Ref.} }
\startdata
CS 22183-015 &5450 3.0   & $-$2.82 & +2.33 & $+$1.85  & 7    & +2.09  &<0.7    & rs & (21) \\
CS 22877-001 & 5100 2.2  & $-$2.72 & +1.0  & $-$0.49 & >10   & 0.0     & 1.20   & no & (9)    \\
CS 22887-048 &6500 3.7   & $-$1.75 & +1.46 & $+$1.68  & 3    & +1.37  &<1.4    & rs & (21) \\
CS 22892-052 & 4710 1.5  & $-$3.1  & +1.05 & $+$0.96 & 15    & +1.0   & 0.15   & r & (1)    \\
             & 4850 1.6  & $-$3.03 & +0.92 & $+$1.01 & 16    & +0.51  & 0.20   & r & (2) \\
CS 22898-027 & 6250 3.7  & $-$2.26 & +2.2  & $+$2.23 & 15    & +0.9   & 2.10   & rs & (3)   \\
             &6000 3.5   & $-$2.44 & +2.05 & $+$2.47  & 12   & +1.41  & 2.18   & rs &  (21) \\
CS 22947-187 &5200 1.5   & $-$2.51 & +1.07 & $+$1.24  & 3    & +1.73  &<0.5    & s & (21) \\
CS 22948-027 & 4600 1.0  & $-$2.57 & +2.00 & $+$1.85 & 10    & +1.80  & <1.0   & rs & (9)    \\
CS 22949-008a &6300 3.5  & $-$2.09 & +1.55 & $+$1.32  &>30   & +0.41  &<1.0    & rs & (21) \\
CS 22949-008b &5300 4.7  & $-$2.09 & +1.55 & $+$1.32  &>30   & +0.41  &<1.0    & s & (21) \\
CS 22949-037 & 4900 1.5  & $-$3.97 & +1.17 & $-$0.58 & 3     & +2.57   & <$-$0.30 & no & (11) \\
CS 22956-028 & 7038 4.3  & $-$1.89 & +1.82 & $+$0.56 & 5     & +1.52  & 2.0    & no & (8)   \\
CS 22958-042 & 6250 3.5  & $-$2.85 & +3.17 & $-$0.53 & 9     & +2.17   & <0.6   & no & (16)   \\
CS 22964-161p & 6050 3.7 & $-$2.39 & +1.49 & $+$1.45 & ...   & ...    & 2.09   & s & (19)  \\
CS 22967-007 & 6479 4.2  & $-$1.79 & +1.82 & $+$2.09 & >60   & +0.92  & <1.6   & s & (8)   \\
CS 29495-042 & 5544 3.4  & $-$1.86 & +1.32 & $+$1.83 & 7     & +1.32  & <1.0   & s & (8)   \\
CS 29497-030 & 6650 3.5  & $-$2.70 & +2.38 & $+$2.17 & >10   & +1.88  & <1.10  & rs & (4)    \\
CS 29497-034 & 4800 1.8  & $-$2.90 & +2.63 & $+$2.03 & 12    & +2.38  & 0.10   & rs & (10)  \\
             & 4983 2.1  & $-$2.55 & +2.42 & $+$1.79 & 20    & +2.32  & <0.10  & s & (8)   \\
CS 29502-092 & 5000 2.1  & $-$2.76 & +1.0  & $-$0.82 & 20    & +0.7    & <1.2   & no & (9)    \\
CS 29512-073 &5600 3.4   & $-$2.04 & +1.2  & $+$1.12  &>60   & +0.56  & 1.93   & s & (21) \\
CS 29526-110 & 6800 4.1  & $-$2.06 & +2.07 & $+$2.39 & ...   & ...    & <2.3   & rs & (5)    \\
CS 29528-041 & 6150 4.0  & $-$3.30 & +1.61 & $+$0.97 & ...   & +3.09  & 1.71   & rs & (16)\\
CS 30315-091 & 5536 3.4  & $-$1.66 & +1.32 & $+$1.62 & >60   & +0.42  & <1.0   & s & (8)   \\
CS 30322-023 &4100 -0.3  & $-$3.28 & +0.49 & $+$0.41  & 4    & +2.7   &<$-$0.3   & low-s & (21) \\
CS 30323-107 & 6126 4.4  & $-$1.73 & +1.12 & $+$1.90 & 9     & +0.82  & <1.1   & s & (8)   \\
CS 30338-089 & 5202 2.6  & $-$1.73 & +1.52 & $+$1.87 & 8     & +0.82  & <0.5   & rs & (8)   \\
CS 31080-095 & 6050 4.5  & $-$2.85 & +2.71 & $+$0.77 & >40   & +0.72  & 1.73   & no & (16)   \\
G 77-61      & 4000 5.0  & $-$4.03 & +2.49 & $<+$1.00& 5     & +2.48  & <1.0   & no & (14)    \\
HD 198269    &4500 1.0   & $-$1.69 & +0.75 & $+$1.12  & 5    & +1.11  &<0.2    & s & (21)  \\
HE 0024-2523 & 6625 4.3  & $-$2.70 & +2.62 & $+$1.52 & 6     & +2.12  & <1.50  & s & (7)    \\
HE 0557-4840 & 4900 2.2  & $-$4.75 & +1.65 & $+$0.03 & ...   & ...    & <0.7   & no & (17)  \\
HE 0107-5240 & 5100 2.2  & $-$5.39 & +3.70 & $<+$0.78 & >60  & +2.43  & <1.12  & no &(13)   \\
HE 1327-2326 & 6180 4.5  & $-$5.65 & +3.90 & $<+$1.42 & ...  & +4.23  & <1.5   & no & (12)  \\
             & 6180 3.7  & $-$5.66 & +4.26 & $<+$1.66 & ...  & +4.71  & <1.5   & no & (12)   \\
HE 1410-0004 & 4985 2.0  & $-$3.09 & +2.09 & $+$1.13 & ...   & ...    & <1.32  & s & (15) \\
HKII17435-00532&5200 2.15& $-$2.23 & +0.68 & $+$0.86 & ...   & ...    & 2.16   & low-s & (18)    \\
LP 625-44    & 5500 2.8  & $-$2.71 & +2.1  & $+$2.74 & 20    & +1.0   & 0.40   & rs & (6)    \\
LP 706-7     & 6250 4.5  & $-$2.55 & +2.1  & $+$1.98 & 15    & +1.2   & 2.25   & rs & (3) \\
             & 6200 4.3  & $-$2.53 & +2.14 & $+$2.08 & ...   & ...    & 2.3    & rs &  (5)    \\
SDSS 0036-10 & 6500 4.5  & $-$2.41 & +2.32 & $+$0.29 & ...   & ...    & <2.0   & no & (5)  \\
SDSS 0126+06 & 6600 4.1  & $-$3.11 & +2.92 & $+$2.75 & ...   & ...    & <2.2   & rs & (5)  \\
SDSS 0924+40 & 6200 4.0  & $-$2.51 & +2.72 & $+$1.81 & ...   & ...    & <2.0   & s & (5)  \\
SDSS 1707+58 & 6700 4.2  & $-$2.52 & +2.10 & $+$3.40 & ...   & ...    & <2.5   & rs & (5)  \\
SDSS 2047+00 & 6600 4.5  & $-$2.05 & +2.00 & $+$1.50 & ...   & ...    & <2.3   & s & (5)  \\
SDSS J1349-0229 &6200 4.0& $-$3.0  & +2.82 & $+$2.17  &  >30 & +1.60  & <1.5   & rs &(20)\\
SDSS J0912+0216 &6500 4.5& $-$2.5  & +2.17 & $+$1.49  &  >10 & +1.75  & <1.4   & rs &(20)\\
SDSS J1036+1212 &6000 4.0& $-$3.2  & +1.47 & $+$1.17  &  ... & +1.29  & 2.21   & rs &(20)  
\enddata
\tablerefs{
(1) \citet{Sneden2003}, (2) \citet{Cayrel2004}, (3) \citet{Aoki2002b}, (4) \citet{Sivarani2004}, (5) \citet{Aoki2008}, (6) \citet{Aoki2001}, (7) \citet{Lucatello2003}, (8) \citet{LucatelloPhD},
(9) \citet{Aoki2002c}, (10) \citet{Barbuy2005}, (11) \citet{Depagne2002}, (12) \citet{Aoki2006}, (13) \citet{Christlieb2004}, (14) \citet{Plez2005},
(15) \citet{Cohen2006}, (16) \citet{Sivarani2006}, (17) \citet{Norris2007}, (18)
\citet{Roederer2008}, (19) \citet{Thompson2008}, (19) \citet{Behara2010}, (21) This work}
\end{deluxetable}
}

\begin{figure}[h!]
\includegraphics[width=12cm,angle=-90]{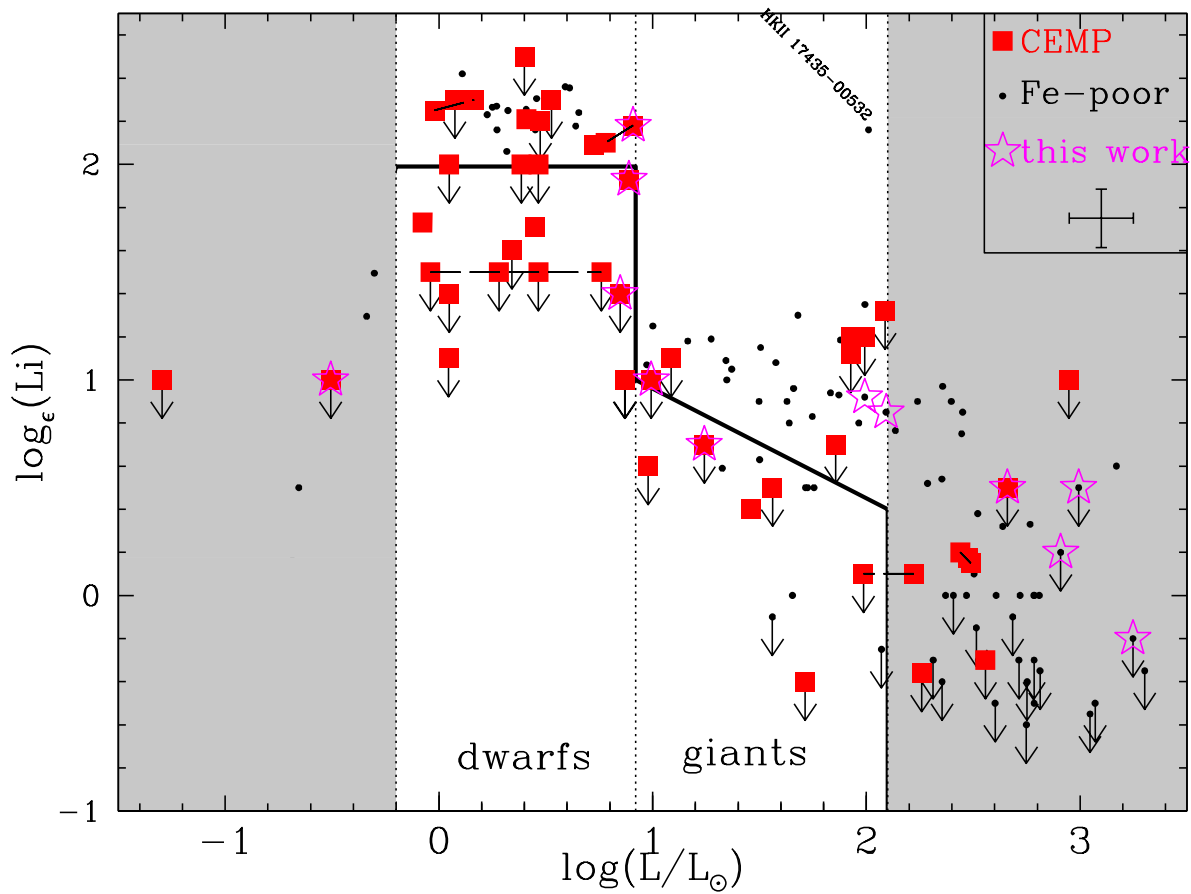}
\caption{Li abundance as a function of luminosity in metal-poor stars. 
The solid line represents the
empirical limit between Li-normal and Li-depleted CEMP stars. Please the
text for a discussion of the expected Li trends in metal-poor stars. 
Only the ``dwarfs'' and the ``giants'' will be discussed in the following
figures (i.e., all the stars in the shaded area will not be represented or
discussed). There is a broad range of Li abundances for CEMP stars. The 
Li abundance in the Li-normal CEMP stars is fully compatible with the Li abundances
of metal-poor stars without carbon-enhancement and, in particular, 
no examples of stars with Li abundances above the Spite plateau are observed.}
\label{fig:LivsL}
\end{figure}

In Figure~\ref{fig:LivsL}, we plot the Li abundance as a function of
luminosity for the extended sample of CEMP stars. We also plot some non-C-rich
metal-poor stars ([Fe/H]$<-1.5$) extracted from the SAGA database
\citep{Suda2008}. These non-C-rich metal-poor stars are used as ``standard''
stars to define empirically the limit where Li can be considered as abnormally
depleted. As expected from stellar evolution, the Li abundance should be constant
during the main sequence. Concerning the giant phase, the occurence of the first dredge-up
 deeply mixes the Li-rich material of the surface with
Li-depleted material. This is indeed seen in Fig.~\ref{fig:LivsL}.
However, the observed limit for the drop of Li abundance due to the first dredge up appears at $\log(L/L_\odot) \sim 1.0$ rather than $\log(L/L_\odot) \sim 0.7$ as expected by the models.
Moreover, the least squares fit of the non-CEMP data shows a
slight trend from ($\log(L/L_\odot) \sim 1.0$) up to $\log(L/L_\odot) = 2.1$. 
This trend is not expected by models.
Nevertheless, we kept the observation of non-CEMP stars as
a reference for the Li-depleted line such that it is defined by a constant value corresponding to the Spite-plateau value minus a
standard error in the Li abundance measurement (i.e. $\log_\epsilon(Li)=$2.0) for the dwarfs, and by the least
squares fit of these data minus a standard error in the Li abundance measurement for the giants. We
disregard the stars with $\log(L/L_\odot) < -0.2$ (and with $\log(L/L_\odot) > 2.1$) because
they all have substantially destroyed their Li -- in keeping with the standard
evolution of non-C-rich stars -- and thus do not provide any useful information.

In this figure, we first notice that CEMP stars exhibit a broad range of Li
abundances for their evolutionary stage. To discuss the origin of the observed
scatter in Li abundances more quantitatively in the following sections, we respectively
refer to ``Li-normal'' stars and ``Li-depleted'' stars, the stars above and the
stars below the thick line in Figure~\ref{fig:LivsL}, respectively. Logically, only the stars
with only an upper limit on their Li abundance below the depleted line can be
classified as Li-depleted stars, but none above this line can be
classified with certainty. Nevertheless, all the stars with only an upper limit on
their Li abundance are always represented by the same ``v'' sign.

There is one Li-rich star in Fig. \ref{fig:LivsL}, HKII 17435-00532
\citep{Roederer2008}, whose unusual Li abundance has been interpreted as the result of
self-pollution from extra mixing on the RGB or early AGB; we will not
consider this star any further in our discussion.

\subsection{Discussion}
\subsubsection{The Origin of the Li-normal Stars}\label{sec:linormal}

If the AGB mass-transfer model is correct, then both the large C abundances and
the normal Li abundances need to be the result of transferred AGB material
mixing with the material in the convective envelope of the CEMP star. Initially,
before mass transfer took place, the star that is now observed as a CEMP star
very likely had the same composition as any extremely metal-poor star, namely no
C-enhancement and normal Li abundance. As more and more AGB material is
transferred, and the ratio of mass in the convective envelope of the secondary
component to transferred mass decreases, the C/H ratio on the surface of the
CEMP star will increase. According
to the yields we have adopted for this study, in most cases,
the Li abundance in the AGB star is lower than the Spite plateau value. Consequently,
the more AGB material is transferred, the more the Li/H ratio will
drop. In a few specific cases depending on the model assumptions, the
abundance of Li in the AGB star is higher than the Spite plateau value. In these cases,
the more AGB material is transferred, the more the Li/H ratio will
increase, along with C. In any case, the predicted C, N, and Li abundances of
the CEMP star depends on the amount of dilution of the AGB material, the
AGB yields, and the original abundances in the CEMP star.

\begin{figure}[h!]
\includegraphics[width=12cm,angle=-90]{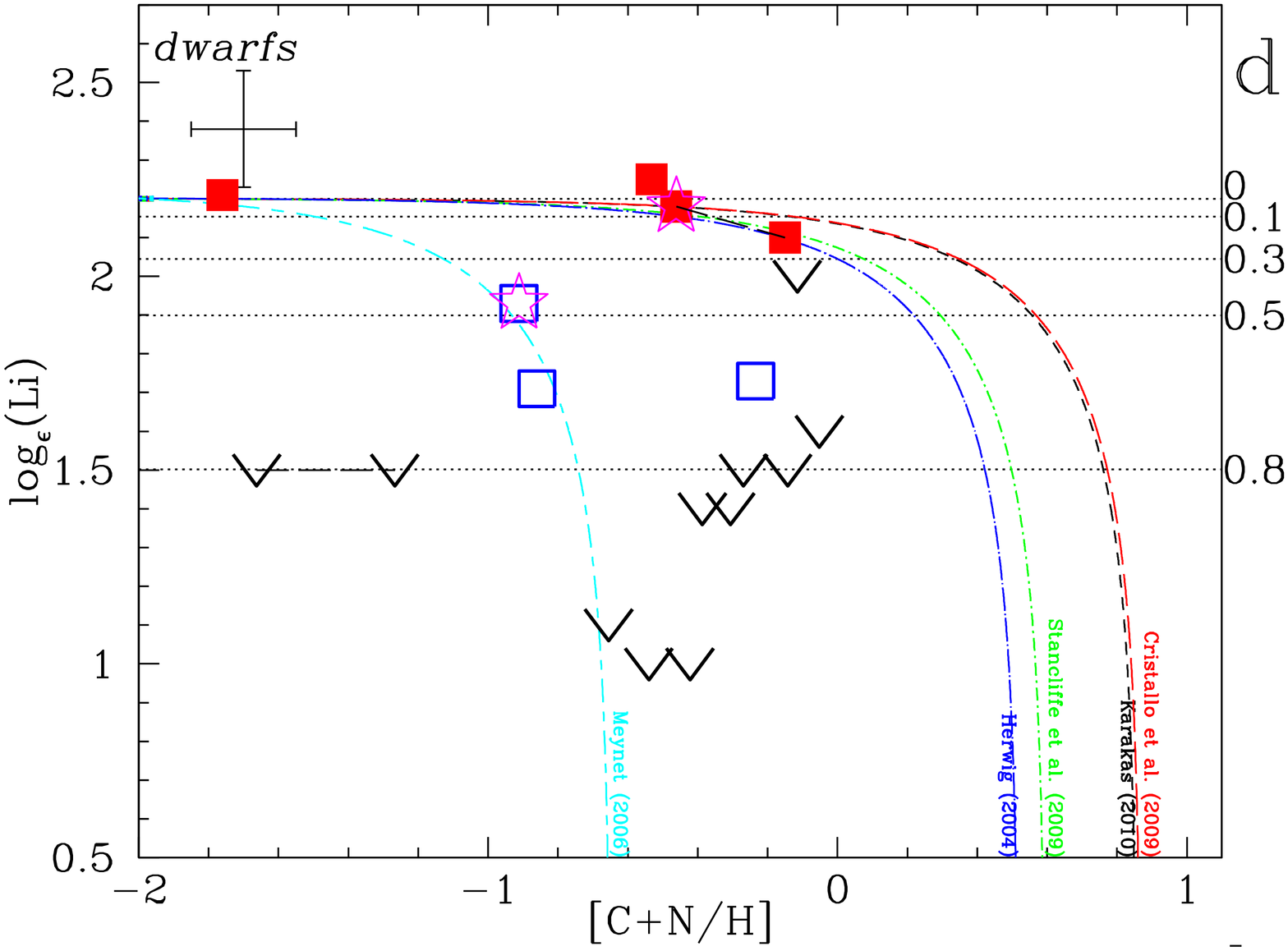}
\caption{Lithium abundance as a function of [(C+N)/H] in CEMP dwarfs. 
The filled red squares indicate Li-normal stars (i.e. $\log_\epsilon(Li)
\geqslant 2.0$), open blue squares indicate Li-depleted stars (i.e.
$\log_\epsilon(Li) < 2.0$), and the v-signs indicate upper limits on Li. 
Values for a star analyzed by more than one study are connected by a line.  The magenta
stars represent the stars analyzed in this paper. The different lines represent
the dilution of the yields of metal-poor 2\,M$_\odot$ AGB models
\citep{Karakas2010,Stancliffe2009Li, Herwig2004,Cristallo2009} and of 60\,
M$_\odot$, rapidly rotating massive stars \citep{Meynet2006} with material that
has $\log_\epsilon(Li)=2.2$, $\log_\epsilon(Li)=-1.0$ and $[(C+N)/H]=-2$ (value
taken form the observation of metal-poor stars). The dotted lines indicate
various dilution factors. Only dwarfs are plotted, as the initial Li abundance before any
processing by the CEMP star can be set confidently to the Spite plateau. The Li-normal
CEMP stars are consistent with a dilution scenario. Note also that all the AGB
models are sufficiently consistent with each other for the purposes of this
discussion.}
\label{fig:LivsC+N_dilu_models}
\end{figure}

To test if any observed combination of Li, C, and N can be explained by AGB mass
transfer, we compare the measurements in the extended sample with different
amounts of dilution of material having 2\,M$_\odot$ AGB yields for different
models with different assumptions (Fig.~\ref{fig:LivsC+N_dilu_models}). We chose
2\,M$_\odot$ because this is the only value that permits comparison between all
models available in the literature (the influence of the mass is discussed
next).  We define the dilution factor as the fraction of mass transferred from the AGB in the total mass of the
convective envelope of the companion after the transfer. We plot
[(C+N)/H], rather than [C/H], because the amount of C that is converted into N in
AGB stars is currently not well-understood \citep[see, e.g., ][]{Johnson2007}. We
also choose to plot [(C+N)/H], rather than [(C+N)/Fe], because C is a primary
element in AGB stars. There is not a strong dependence on metallicity for C
production in AGB stars for the metallicity range we 
considered \citep[e.g., the yields of a 3\,M$_\odot$ AGB of] [give
{[}(C+N)/H{]}=+0.97 for Z=4$\times$10$^{-3}$ and {[}(C+N)/H{]}=+1.17 for Z=10$^{-4}$]
{Karakas2007}, thus the comparison between CEMP stars of different metallicities
and models of a single metallicity (Z=10$^{-4}$) is valid. Note that we
ignored the effect of possible hydrogen injection episodes that are predicted
to occur in stars with
low metallicity \citep{Campbell2008,Lau2009,Suda2010}. The main results
of H-ingestion episodes is large enhancements of
C and N at the stellar surface.

We computed dilution tracks using the relation:
{\small
\begin{eqnarray}
[(C+N)/H] & = & \log\Big( (A(C)_{AGB}+A(N)_{AGB}) \times d +  (A(C)_{init}+A(N)_{init})\times (1-d)\Big) - \log\Big(A(C_\odot)+A(N_\odot)\Big) \\
\log_\epsilon(Li) & = & \log(10^{\log_\epsilon(Li)_{AGB}} \times d + 
10^{\log_\epsilon(Li)_{init}} \times (1-d)) 
\end{eqnarray}}
where $d$ is the dilution factor, varying between 0 (no mass transfer) and
1 (convective envelope of the CEMP star comprised entirely of AGB-transferred material).
The initial C+N yields ($\log_\epsilon(C+N)_{AGB}$) are taken from various
models. We assume an almost Li-free accreted material
$(\log_\epsilon(Li)_{AGB}=-1.0)$, as well as an initial Spite plateau Li
($\log_\epsilon(Li)_{init}=2.2$) and a $[(C+N)/H]_{init}=-2.0$, compatible with
non-C-rich main-sequence metal-poor stars with [Fe/H] = $-2.0$
\citep[e.g., ][]{Asplund2006}. 

\begin{figure}[h!]
\includegraphics[width=12cm,angle=-90]{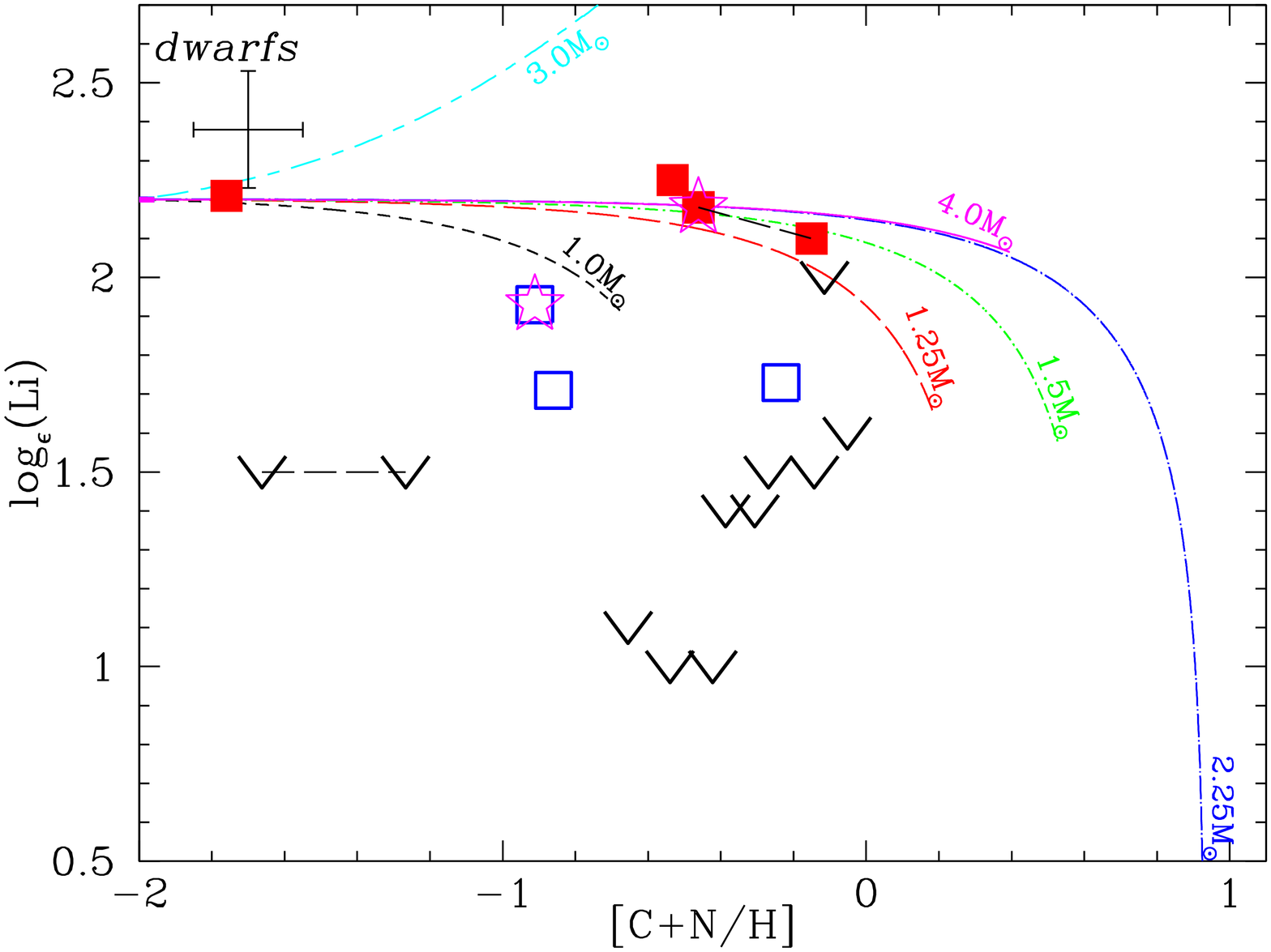}
\caption{Same as Fig.~\ref{fig:LivsC+N_dilu_models}, but with AGB yields of
C, N, and Li of different 
masses from \citet{Karakas2010}. The dilution tracks are able to match the
highest C enhancement but do not match the low-C and low-Li CEMP stars. Note that
the final log$_\epsilon$ (Li) of the 3M$_\odot$ track reaches Li-rich values.}
\label{fig:LivsC+N_dilu_masses}
\end{figure}

Figure~\ref{fig:LivsC+N_dilu_models} shows that there is a regime where C
remains high, while the Li abundance is almost identical to the Spite Li
plateau. This corresponds to a dilution factor of $\sim$ 10\%, similar
to that found by \citet{Thompson2008}. 

However,
not all AGB stars have masses of 2\,M$_\odot$, and in
Fig.~\ref{fig:LivsC+N_dilu_masses}, we plot the dilution tracks for yields from
AGB stars with a range of masses from \citet{Karakas2010}, including Li yields.
Fig.~\ref{fig:LivsC+N_dilu_models} and Fig.~\ref{fig:LivsC+N_dilu_masses} also
show that the observed abundances of Li-normal stars fall into this regime,
suggesting that the Li abundances observed in these stars might not require any production
by the AGB companion, in contrast to the conclusions of
\citet{Norris1997Cstars}, \citet{Sivarani2006}, and
\citet{Thompson2008}.
 
We also see in this figure that it is the {\it Li-depleted dwarfs} with low
[C/H] that cannot be reproduced by the dilution tracks. Even the 
lowest-mass AGB star shown in Fig.~\ref{fig:LivsC+N_dilu_masses} has a sufficiently high Li yield
that it cannot reproduce the Li-depleted dwarfs. 
Fig~\ref{fig:LivsC+N_dilu_masses} illustrates another prediction of the AGB
models that makes the creation of Li-depleted dwarfs more difficult. The Li
yields of the Karakas (2010) AGB models are not zero for stars of any mass, and
for each mass, the dilution curves stop at the log(Li) abundance of pure AGB
material. However, there are uncertainties of Li yields from the
AGB models due to mass loss, for example. While taking the yields from an AGB model and then applying
dilution factor is a simple way to estimate the abundance of the
companion, it does not exactly reproduce what the composition of the
mass transferred in a real binary system. Once the mass transfer
starts, the envelope of the AGB primary may be lost and the evolution
truncated \citep{Hurley2002,Izzard2009}. Particularly for Li, it is very likely that the surface
abundances of Li varies by more than one order of magnitude from early
AGB till the end of evolution \citet{KarakasPhD}. Therefore, the Li abundance of the
companion depends on when mass transfer starts and hence the period of
the binary system.  
However, when comparing Fig.~\ref{fig:LivsC+N_dilu_models} and \ref{fig:LivsC+N_dilu_masses}, the dilution tracks do not much depend on the AGB Li yields. For instance, if more Li depletion is considered, the dilution tracks tends to extend towards lower Li values but still do not encompass the CEMP data. Therefore, within the assumption that [C+N] yields of single AGB star are valid for CEMP binary systems, the presence of Li-depleted CEMP stars cannot be explained.

\subsubsubsection{Comparison with Ba abundances and \twch/\thch{} ratios}\label{sec:Ba_C13vsLi}

One alternative explanation for the results in the previous section is that
current AGB models predict too high C+N yields. \citet{Stancliffe2010} pointed
out that, to fit the mild C abundances of the SB2 CS~22964-161, they need either
to invoke some mixing process or to decrease the C yields of the former AGB
companion. Additionally, \citet{Masseron2010} showed that the [Ba/C] measured in
CEMP stars is too high compared to the predictions, highlighting a problem for
the predictions of either the Ba or the C production. To explore this
possibility, we checked whether the dilution models agree with other observations
of CEMP stars, in particular the Ba abundances, which are sensitive to
$s$-process production, and the \twch/\thch{} isotope ratios, which are sensitive to
proton-capture reactions that happen under the same physical conditions as the
Cameron-Fowler mechanism.
 
\begin{figure}[h!]
\includegraphics[width=12cm,angle=-90]{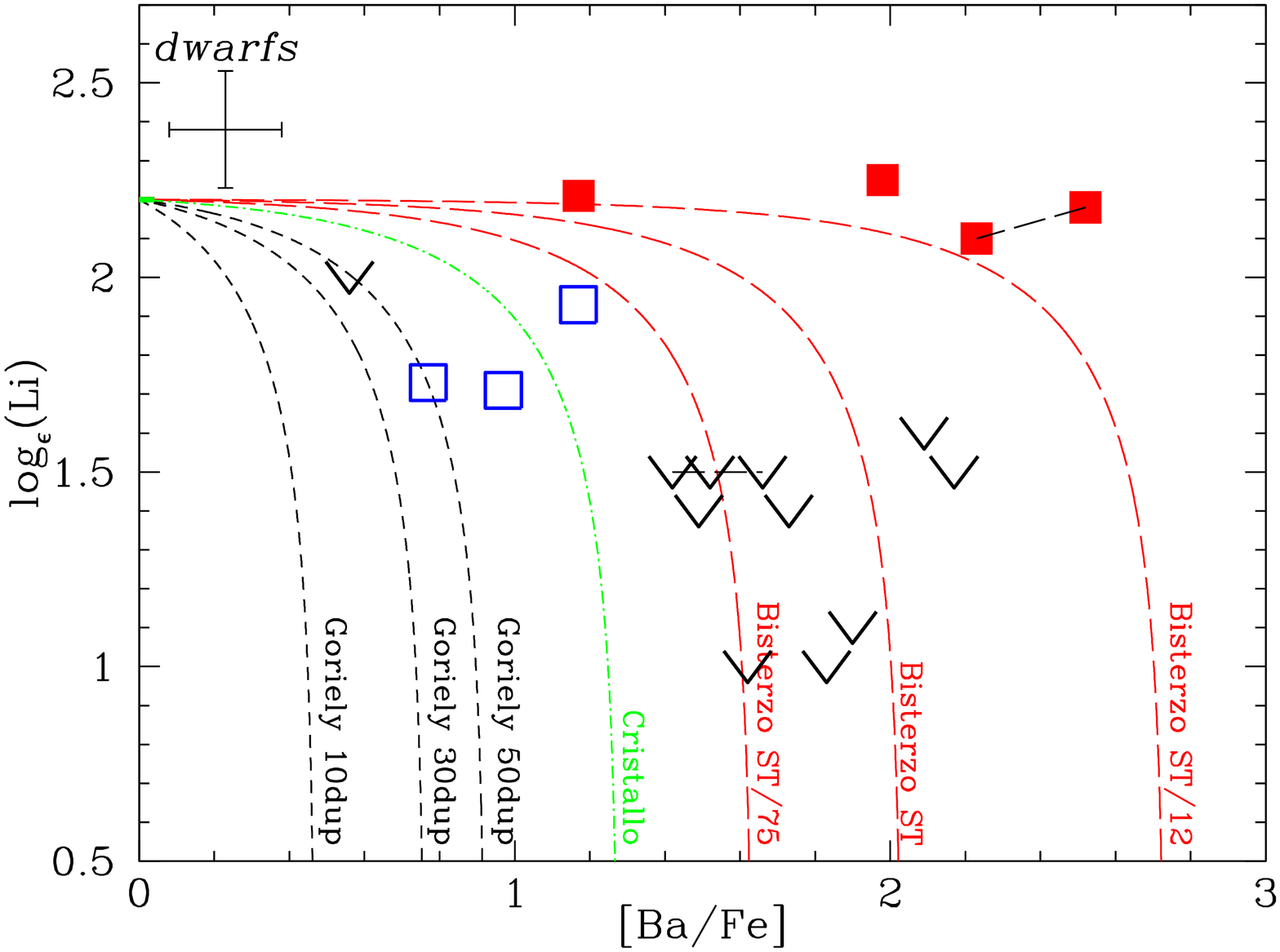}
\caption{Same as Fig.~\ref{fig:LivsC+N_dilu_models}, but for the abundance of Li as a function of [Ba/Fe]. Theoretical yields are extracted from Goriely (private communication) for 1.5\,M$_\odot$, Z=10$^{-4}$, (10, 30, and 50 dredge ups), from \citet{Cristallo2009} for 2\,M$_\odot$, Z=10$^{-4}$, and from \citet{Bisterzo2010} for 1.5\,M$_\odot$, Z=10$^{-4}$, standard, standard/12 and standard/75 $^{13}$C pocket sizes.}
\label{fig:LivsBaFe}
\end{figure}

Figure~\ref{fig:LivsBaFe} shows the abundance of Li as a function of [Ba/Fe]
compared with dilution tracks, similar to those obtained for C+N. In this case
we use [Ba/Fe]=0 as the secondary's initial Ba abundance. 
Here the dilution tracks also cannot reproduce the Li-normal
and the Li-depleted stars based on a single set of models, as the
full range is spanned only if we adopt different sets of models.
However this conclusion is less secure than for C+N because
the available models cannot predict the abundances of
Ba and other s-process elements in CEMP without {\it requiring}
an ad-hoc parametric mixing mechanism to form a $^{13}C$ pocket and 
facilitate s-process production. 
Note that, while the observed data in the figure spans a wide metallicity range, models shown have
[Fe/H]=-2.3 only. However, Ba production in metal-poor AGB stars depends proportionately on the metallicity \citep{Straniero1995,Goriely2000} and thus [Ba/Fe] is almost constant for a given AGB mass. This fact has been confirmed by \citet{Masseron2010}, who noticed 
that [Ba/Fe] is strongly correlated to C in CEMP stars. For this reason, assuming 
a metallicity of [Fe/H] = -2.3 for all the models to represent the bulk of the CEMP-s stars 
is suitable for purposes since we are looking for global trends and not 
trying to fit abundance patterns of individual objects.

\begin{figure}[h!]
\includegraphics[width=12cm]{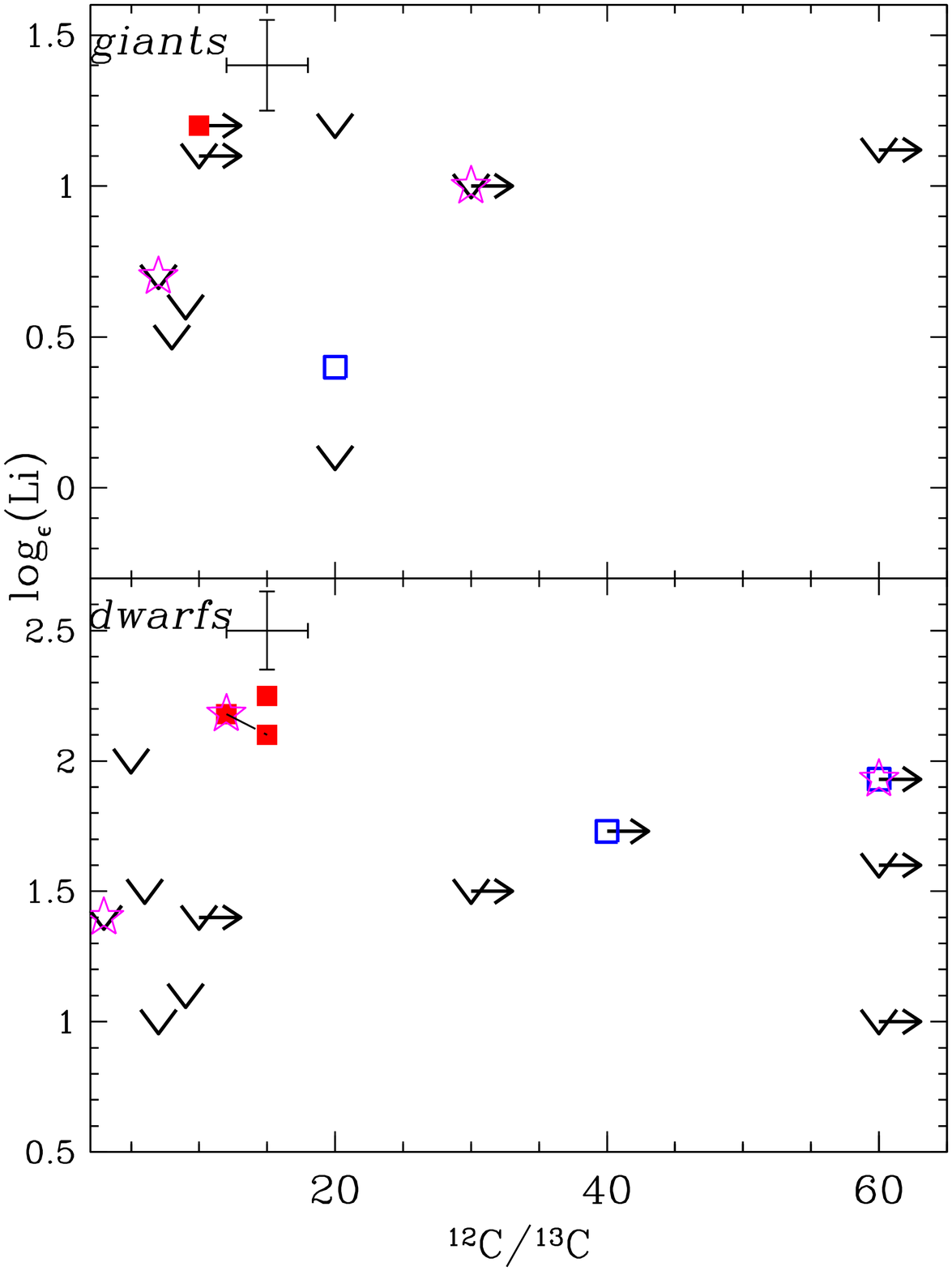}
\caption{Lithium abundances as a function of $\rm ^{12}C/^{13}C$ ratios in 
CEMP ``dwarfs'' (lower panel) and ``giants'' CEMP (upper panel). 
Symbols are identical to Fig.~\ref{fig:LivsC+N_dilu_models}. There does
not seem to be a correlation between Li depletion and  $\rm ^{12}C/^{13}C$ ratio.}
\label{fig:Livs12C13C}
\end{figure}

As shown by \citet{Masseron2010}, CEMP stars
exhibit a correlation between their $\rm ^{12}C/^{13}C$ ratios and C/N. This is clear
evidence that CN cycling occurred in the AGB stars. Complete or partial CN cycling indicates that the envelope of the star has
reached sufficiently high temperatures that Li can be easily destroyed. Naively,
one might expect that the more complete the CN cycle is (hence the lowest $\rm
^{12}C/^{13}C$ ratios are reached), the more Li is destroyed. 
On the other hand, \citet{Masseron2010} demonstrated that extra mixing was at work below the
convective envelope in the companions of CEMP stars. Li production by
extra mixing may be possible \citep{Dominguez2004,Stancliffe2010} 
and is expected to also show a correlation with the $\rm ^{12}C/^{13}C$ ratio. In
both scenarios, a correlation between Li abundances and C isotope ratios would
be expected if the Li/H ratio was mainly determined by the Li/H ratio in the AGB gas. 
However, Fig.~\ref{fig:Livs12C13C} shows that both Li-normal CEMP stars with low
$\rm ^{12}C/^{13}C$ ratios and Li-normal CEMP stars with high $\rm
^{12}C/^{13}C$ ratios exist. 
Li production or destruction is
extremely sensitive to the adopted mixing parameters \citep{Wasserburg1994}.
While continued investigation of the yields and the impact of extra mixing in
AGB stars is important for this discussion, there is no conclusive evidence that
they are uncertain enough to explain the broad dispersion of Li in CEMP stars.
Actually, it is puzzling that CEMP stars with apparently similar companions (at
least with identical extra mixing parameters leading to identical C isotopes and
$s$-process enrichments) can exhibit such a broad variety of Li abundances
(Fig.~\ref{fig:LivsBaFe} and \ref{fig:Livs12C13C}). This may indicate that
the carbon isotope ratios seen in the CEMP stars reflect the nucleosynthesis in the AGB
stars, whereas the Li abundances reflect processing in the CEMP stars themselves.

\subsubsection{The expected distribution of Li abundances}\label{sec:statistics}

We have argued that the Li-normal CEMP stars can naturally appear in the AGB
mass-transfer scenario. However, some AGB stars within a limited mass range are
expected to produce Li in excess of the Spite plateau value and, if such an
enriched gas is transferred to a companion, that companion should appear as a
Li-rich star, unless in all cases processes in the CEMP star destroys the Li. We
find no such Li-rich stars in our VLT sample, nor in
the extended sample. We can test if the lack of Li-rich stars is expected in a
sample of the size discussed in this paper, if AGB mass transfer is responsible
for the creation of CEMP stars. As mentioned in the Introduction, the Li yields
from AGB stars and the amount of mass transferred are uncertain, but we can
calculate the statistics for several cases (see Table \ref{tab:stats}). In one case,
we assume that the percentage of transferred mass compared to convective envelope mass is 10\%
($d=0.1$), similar to the ratio that \citet{Thompson2008} found for
CS~22964-161. In another case, we assume that the dilution factor is so large
($d=1$) that the abundances on the surface of the CEMP star reflect the yields
of the AGB star for all elements. This case reflects the fact that
\citet{Izzard2009} computed that as much as 0.1\,M$_\odot$ should be typically
accreted onto the relatively small (i.e., a few $10^{-3}\,M_\odot$) convective envelope of
the CEMP star. In contrast, we assume in the last case ($d=0.01$) that the
accreted material has been highly diluted into the CEMP star \citep[possibly by
thermohaline mixing, see][]{Stancliffe2007}.

We use the yields of \citet{Karakas2010}, performing a linear interpolation
between the model yields to predict the Li yield for stars of any mass. For AGB
stars with masses less than 1.0\,M$_{\odot}$ we adopt the ratios for a 1.0\,M$_{\odot}$ 
AGB star. We also need to assume the distribution of mass ratios (q)
for binaries and the range of primary and secondary masses that will produce
observable CEMP stars today. 

We consider several cases for the mass range of the primaries: the low-mass limit 
is between 0.83\,M$_\odot$ and 1.5\,M$_\odot$, while the high-mass limit is set
to either 3\,M$_\odot$ or 8\,M$_\odot$. The 0.83\,M$_\odot$ limit for the AGB
stars at the low-mass end comes from the fact that observations of the very
low-mass, low-gravity star CS~30322-023 are compatible with the theoretical
expectation that it is a low-mass AGB which has dredged up C, N, and $s$-process
elements \citep{Masseron2006}. On the other hand, this low-mass limit for third
dredge-up is metallicity- and model-dependent \citep[e.g.,][]{Karakas2002}.
Depending on the assumptions in models, stars with M$<$1.5\,M$_\odot$ are not
expected to produce enough C and $s$-process elements to make CEMP stars, so we
adopt 1.5\,M$_\odot$ as another possible limit. At the high end, the 8\,
M$_\odot$ limit comes from theoretical expectations of the highest mass an AGB
star can have \citep[see for example][]{Ventura2010}. However, in AGB stars of
very low metallicity ($Z=10^{-4}$ with M$>$3\,M$_\odot$), hot-bottom burning
should have occurred, which will produce nitrogen-enhanced metal-poor (NEMP)
stars rather than CEMP stars. 

The limits for the secondaries are always between 0.71\,M$_\odot$ and 0.82\,
M$_\odot$, which come from demanding that the secondary star have a luminosity
where its Li is not expected to be depleted (see Fig.~\ref{fig:LivsL}). We
converted between luminosity and mass using a 12\,Gyr Padua isochrone with
Z=0.0001 \citep{Girardi2002}. For our mass ratios, we followed the lead of
\citet{Izzard2009} and adopted a flat distribution of q values, because there is no
estimate in the literature of q at low metallicity. The initial Li abundance of
the CEMP gas was set to 2.2, thus any secondary with a Li abundance higher
than 2.2 is labeled Li-rich.

\begin{deluxetable}{lccccc}
\tabletypesize{\footnotesize}
\tablewidth{0pt}
\tablecaption{Predicted Distributions of Li Abundances \label{tab:stats}}
\tablehead{\colhead{Primary Masses} & \colhead{Abundance Used} & \colhead{Dilution} & \colhead{\% Li-rich} & \colhead{\% Li-normal} & \colhead{\% Li-poor} \\
\colhead{M$_\odot$} & \colhead{} & \colhead{$d$} & \colhead{log$_\epsilon$ (Li) $> 2.2$}  & \colhead{$2.0 <$ log$_\epsilon$ (Li) $< 2.2$}  & \colhead{log$_\epsilon$ (Li) $< 2.0$}}
\startdata
                           &                         & 0.01 &  4 & 96 &  0 \\
                           &                         & 0.10 &  4 & 96 &  0 \\
0.83\,M$_\odot$-8.0\,M$_\odot$ & K10                 & 0.30 &  4 & 95 &  1 \\
                           &                         & 0.50 &  4 & 73 & 23 \\
                           &                         & 0.80 &  4 &  1 & 95 \\
                           &                         & 1.00 &  4 &  0 & 96 \\
\hline
                           &                         & 0.01 &  2 & 98 &  0 \\
                           &                         & 0.10 &  2 & 98 &  0 \\
0.83\,M$_\odot$-3.0\,M$_\odot$ & K10                 & 0.30 &  2 & 98 &  0 \\
                           &                         & 0.50 &  2 & 76 & 22 \\
                           &                         & 0.80 &  2 &  0 & 98 \\
                           &                         & 1.00 &  2 &  0 & 98 \\
\hline
                           &                         & 0.01 &  3 & 97 &  0\\
                           &                         & 0.10 &  3 & 97 &  0 \\
 1.0\,M$_\odot$-3.0\,M$_\odot$ & K10                 & 0.30 &  3 &  0 & 97 \\
                           &                         & 0.50 &  3 &  0 & 97 \\
                           &                         & 0.80 &  3 &  0 & 97 \\
                           &                         & 1.00 &  3 &  0 & 97  \\

\hline
                           &                         & 0.01 & 10 & 90 &  0 \\
                           &                         & 0.10 & 10 & 90 &  0  \\
1.5\,M$_\odot$-3.0\,M$_\odot$  & K10                 & 0.30 & 10 &  1 & 89 \\
                           &                         & 0.50 & 10 &  1 & 89 \\
                           &                         & 0.80 & 10 &  1 & 89 \\
                           &                         & 1.00 & 10 &  1 & 89 \\
\hline
\hline
                           &                         & 0.01 &  0 & 100 & 0 \\
                           &                         & 0.10 &  0 & 100 & 0 \\
1.0\,M$_\odot$-3.0\,M$_\odot$  & log$_\epsilon$(Li)=$-1$ & 0.30 &  0 &  0 &100 \\
                           &                         & 0.50 &  0 &  0 &100 \\
                           &                         & 0.80 &  0 &  0 &100 \\
                           &                         & 1.00 &  0 & 0 & 100\\
\hline
\enddata
\\
K10: Li AGB yields from \citet{Karakas2010}
\end{deluxetable}

Examination of Table \ref{tab:stats} reveals that, regardless of the
assumptions, less than 10\% of dwarfs should be Li-rich if CEMP stars
are the result of AGB mass transfer. In the dwarf category,
we have 12 stars with a useful Li measurement, none of which are Li-rich.
Therefore, our approximate calculation of the number of Li-rich stars agrees
with observations. We note that any subsequent mixing in the CEMP star that
destroys Li will artificially decrease the number of detected Li-rich stars,
which is another reason why it is not surprising that we have not found any. The
adopted yields also play a role in the expected number of Li-rich stars.
According to the \citet{Ventura2010} yields, for example, Li is produced in
excess of the Spite plateau value for AGB stars with M$>7$\,M$_\odot$, which
should produce substantial amounts of N and form NEMP stars rather than CEMP
stars. We note that in the Galactic disk, the Li abundance in AGB stars that
dredge it up from their interiors can reach very high values (log$\epsilon$(Li)
$\gtrsim 4$), but the majority of the AGB stars ($\sim$90\%) show no Li
enrichment \citep{Abia1993,Uttenthaler2010}. The fact that we find no Li-rich
CEMP stars may indicate that Li production during the AGB phase was rare at the
metallicities of the CEMP stars, although a larger sample is necessary before any
definite statement can be made.

The predicted relative numbers of Li-normal and Li-depleted stars in the
absence of additional depletion in the CEMP stars depend much more on the
amount of dilution than did the number of Li-rich stars. For the cases of $\sim$0-20\%
dilution, the expected number of Li-depleted CEMP stars (i.e., $\log_\epsilon(Li)
<2.0$) should be 0\% and a majority of stars should be Li-normal ($\sim$96\%).
The opposite is true in the case of very large dilution factors (Table
\ref{tab:stats}), indicating the sensitivity of the Li abundance to details
of the mass transfer rather than to AGB nucleosynthesis. We reiterate that
Fig.~\ref{fig:LivsC+N_dilu_models} and Fig.~\ref{fig:LivsC+N_dilu_masses} show that
large dilution factors should be accompanied by large C enhancements, 
therefore at least some fraction of our sample of stars has likely experienced
dilution factors of $\sim$0-20\%. More quantitatively, we can count 15 stars
for which $-1 <[C+N/H] <0 $. This range of C+N approximately corresponds to a
10\% dilution factor. Among these, 12 ($\sim 80\%$) are Li-depleted. This is very
far from $<$10\% expected, based on dilution yet. Yet these stars are Li-depleted,
which must then happen during the lifetime of the CEMP star itself.

\subsubsection{The Origin of the Li-depleted Stars}\label{sec:Lidepleted}

We have shown in Fig.~\ref{fig:LivsC+N_dilu_models} and
Fig.~\ref{fig:LivsC+N_dilu_masses} that the low-C and low-Li stars are difficult to
explain with just a dilution scenario. Therefore, some extra Li depletion should
occur in the atmosphere of CEMP stars. For the case of the CEMP stars, two main
physical mechanisms have been proposed in the literature that could be
responsible for Li depletion: thermohaline mixing and rotation. In this section,
we will examine whether either scenario is compatible with the observations of
CEMP stars.

\subsubsubsection{Thermohaline Mixing}
\begin{figure}[h!]
\includegraphics[width=12cm]{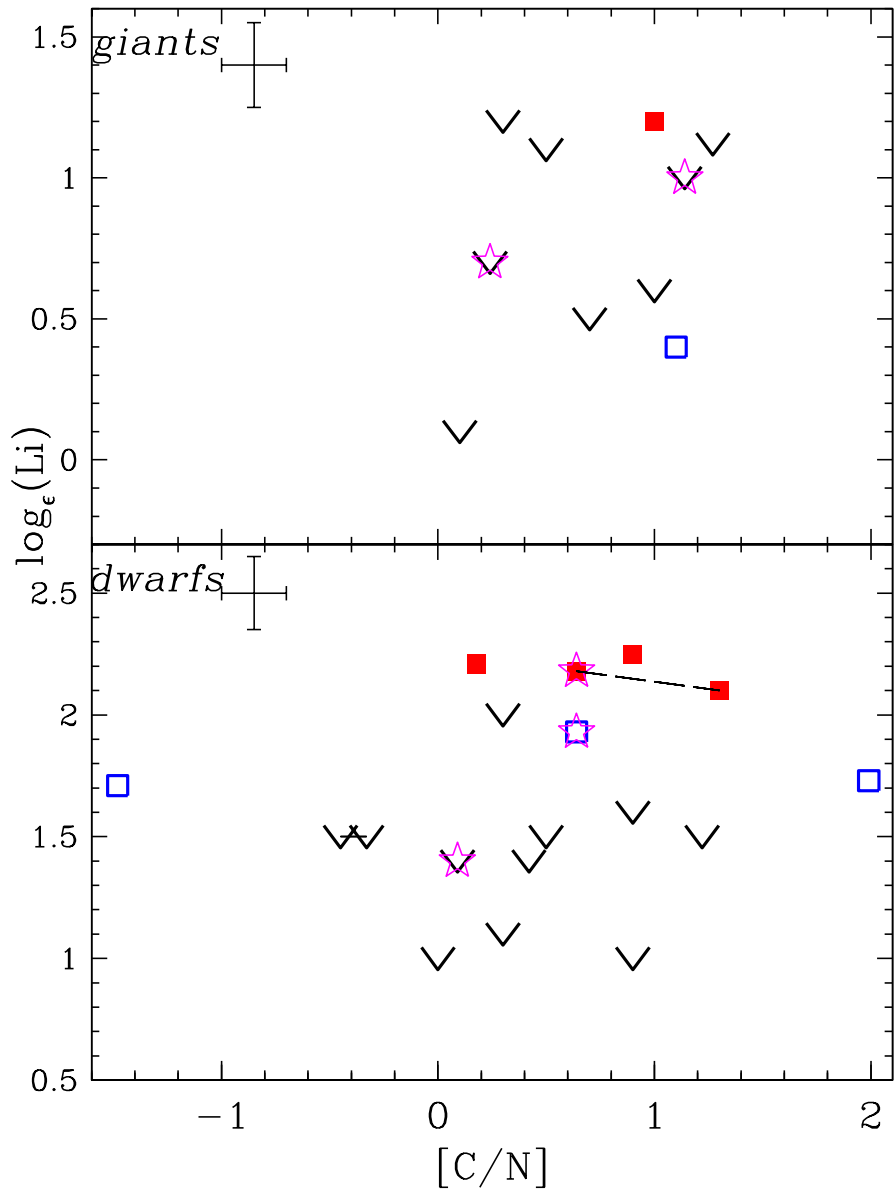}
\caption{Same as Fig.~\ref{fig:Livs12C13C}, but for Li abundance as a function of [C/N].}
\label{fig:LivsCN}
\end{figure}

Because CEMP stars have accreted large amounts of material
from their companions, there are probably changes in their stellar structures
that result. 
Whether thermohaline mixing, caused by the addition of material with a higher
molecular weight on top of a material with lower molecular weight, is sufficient
to explain abundance correlations in stars is still the subject of debate
\citep{Charbonnel2007,Eggleton2008,Denissenkov2010, Palmerini2011}, as it is
dependent on the free parameter that is supposed to represent the thermohaline
``finger'' length. Note that while \citet{Masseron2010} did not find strong evidence
for deep thermohaline mixing in CEMP stars, they could not rule it out. If
it occurs, thermohaline mixing will affect other abundances besides Li.
\citet{Stancliffe2009} predicted that thermohaline mixing will affect the C
isotope ratios in the material transferred from the AGB star. Because of mixing in the
CEMP star, the original AGB composition is not preserved, and there should be a
change in the $\rm ^{12}C/^{13}C$ ratio as a function of the CEMP luminosity
when the stars evolve from dwarfs to giants. If we advance the hypothesis that
the observed Li-depletion spread is related to different thermohaline mixing
efficiency, we would expect that the $\rm ^{12}C/^{13}C$ ratios would reflect
the same efficiency. Fig.~\ref{fig:Livs12C13C} and Fig.~\ref{fig:LivsCN} show no
evidence to corroborate this idea. However, \citet{Stancliffe2008} noted that
gravitational settling could weaken the thermohaline process. Also, Li burns at
2.5$\times$10$^6$~K, while CN cycling starts only at 90$\times$10$^6$~K, so there
is still some room for Li to be destroyed in the CEMP star while $\rm
^{12}C/^{13}C$ remains unchanged from the AGB yields. Another test is that the
thermohaline strength should be correlated with the amount of accreted material,
i.e., the more C is accreted on the surface of the CEMP stars, the deeper the
mixing under the atmosphere of the star, and the larger the Li depletion.
However, the more efficient the thermohaline mixing is, the more the accreted
material is diluted. Hence, it is difficult to confidently predict an
anti-correlation between C+N and Li. 

Nevertheless, we stress that \citet{Stancliffe2009Li} shows that thermohaline
mixing inevitably leads to Li depletion, and thus the presence of Li-normal stars
along Li-depleted stars with the same C+N abundances requires Li production in the
AGB companion, which is uncertain (see Sec.~\ref{sec:Ba_C13vsLi}). Although the
$\mu$-gradient established by the transfer of material should initiate thermohaline
mixing in the CEMP stars' atmospheres, its efficiency and therefore its impact
are difficult to detect, and it might not be the major source of the Li
depletion.
 
\subsubsubsection{Rotation}

An alternative explanation for Li depletion is rotation, where the additional
meridional circulation and shear turbulence drag down some Li to hotter
temperatures and partially destroy it. 
Lithium depletion in dwarfs is not only seen in CEMP stars.
\citet{Norris1997Lidepleted} found three non-C-rich metal-poor stars with
depleted Li. They argued that these stars were likely in binary systems that
somehow caused their Li depletion. \citet{Ryan2001} and
\citet{Ryan2002} further analyzed some of these Li-depleted main sequence halo
stars and presented evidence that rotation has indeed played a role in the Li
destruction. Most of CEMP stars are demonstrated to be old binary systems \citep{Lucatello2005bin}. 
Indeed, increased rotational velocities of CEMP stars are all the more likely.

\begin{figure}[h!]
\includegraphics[width=12cm,angle=-90]{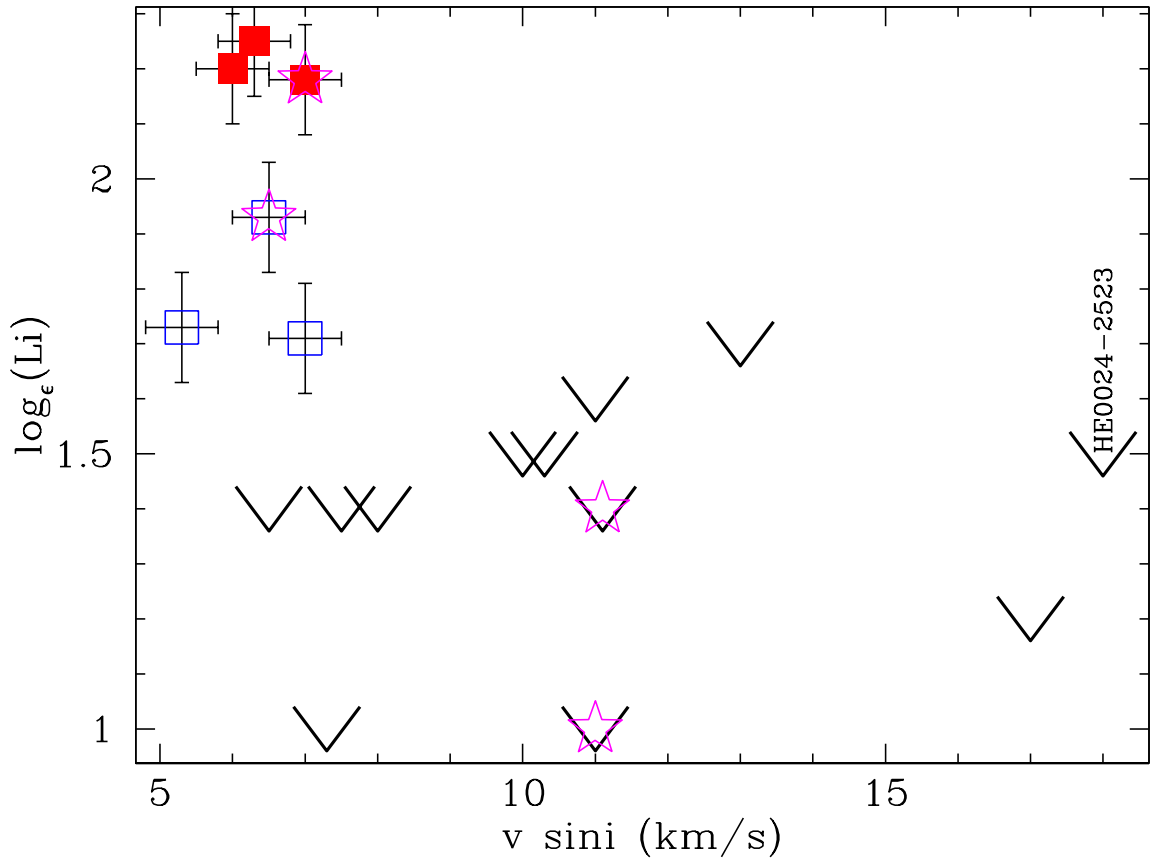}
\caption{Lithium abundance in CEMP dwarfs as a function of rotational
velocities, deduced from line broadening (symbols are the same as in
Fig.~~\ref{fig:LivsC+N_dilu_models}). The stars that exhibit a large broadening
profile are Li-depleted, while the Li-normal stars show small broadening.}
\label{fig:Livsconvol}
\end{figure}

Fig.~\ref{fig:Livsconvol} shows the Li abundances as a function of line broadening
profile for the CEMP dwarfs. Broadening profiles have been derived by setting a
Gaussian profile of synthetic Fe lines to match the observations. For
stars that do not belong to our sample, the synthesis has been made according
to the published stellar parameters and the observations have been obtained
from available archives. By using this technique, the resulting broadening
profile represents either the macro-turbulence velocities in the atmosphere of
the star, its rotation, or both (note that the broadening due to the instrument
should not be a limiting factor in our case because the spectra have a resolving
power high enough to resolve the intrinsic line profiles). {Concerning  macroturbulence
broadening, 3D modeling predicts a broadening of typically 3km/s for metal-poor stars
\citep[e.g., ][]{Asplund2006,Steffen2009}, which is negligible compared to the broadening profiles displayed in Fig.~\ref{fig:Livsconvol}.

Finally, a large fraction of stars have a 
broadening profile $>$ 8 km/s, which cannot be consistent with 
the slow rotational velocity expected for the old population of dwarfs \citep[typically 2km/s,][]{Skumanich1972,Lucatello2003rot}. 
Therefore, we assume that the derived velocities are
mainly due to the rotation of the star, in particular for the highest values.}
It is notable that the largest broadening profile obtained corresponds to HE~0024-2523
\citep{Lucatello2003}, the shortest period binary known among CEMP stars, hence
possibly the fastest rotating star we have considered. \\

In Fig.~\ref{fig:Livsconvol}, all stars with large rotational
velocities show depleted Li. Additionally, all Li-normal stars have
low rotational velocities.  This indicates that rotation is
responsible for depletion in some CEMP stars (at least for the fastest
rotators). For this explanation to work, the degree of mixing
caused by rotation needs to be large enough to destroy without Li
without affecting CN or carbon isotopes, because we see no correlation
between [C/N] and C isotopes and rotation. However, unlike thermohaline
mixing models, models of rotational mixing, at least for sun-like stars,
suggest that it destroys Li without affecting C or N until the subgiant/giant
branch \citep{Pinsonneault1989}.

We note also that Li-depleted CEMP
stars with low rotational velocities exist as well. Nevertheless, due to the
unknown projection angle of the rotation axis of the stars, it is impossible to
conclude for these stars whether the rotation is actually low in these stars or
another mechanism is at work to deplete the Li. It is therefore difficult, with
the current data, to exclude thermohaline mixing as the driver for Li depletion
in CEMP stars. \citet{Charbonnel2010} studied the complex interplay
between rotation and thermohaline mixing and found that rotation favors the
occurrence of thermohaline mixing in low-mass solar metallicity RGB stars.
Furthermore, \citet{Stancliffe2008} show that, beyond rotation and thermohaline
mixing, other mechanisms such as gravitational settling may play some role in
dwarf CEMP stars and subsequently affect the observed Li abundances. In the end, the
various contributions of all these physical processes may explain the observed
Li abundance scatter.

\subsubsection{Li in CEMP Star Subclasses}

\begin{figure}[h!]
\includegraphics[width=12cm]{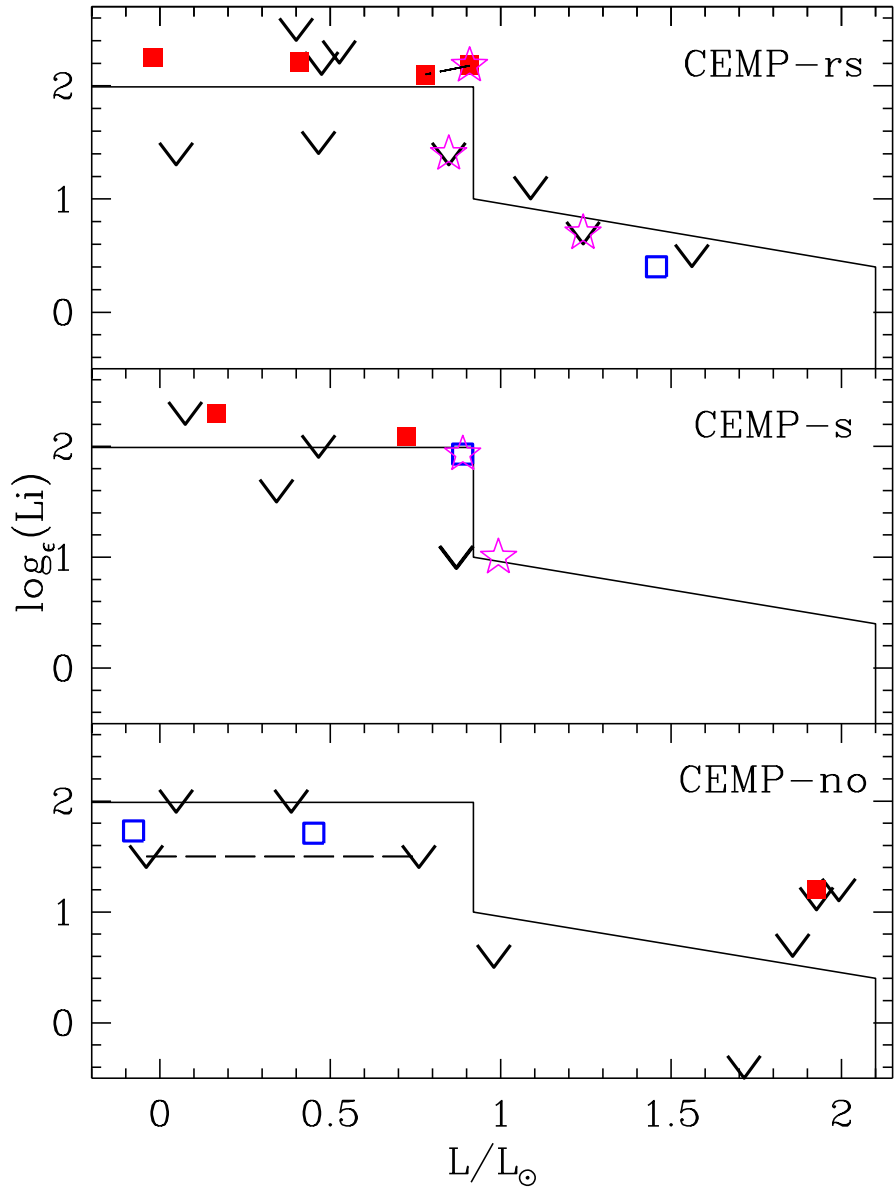}
\caption{Lithium abundance as a function of luminosity for different CEMP subclasses
\citep[for a detailed definition see ][]{Masseron2010}. Symbols are identical
to Fig.~\ref{fig:LivsC+N_dilu_models}. The line shows our adopted separation between
Li-normal and Li-depleted stars. The CEMP-s and CEMP-rs categories encompass
both Li-rich and Li-depleted stars, but the CEMP-no class has only Li-depleted
stars.}
\label{fig:Livssubclasses}
\end{figure}

In view of the ongoing debate about whether the different CEMP subclasses are
the result of the same basic phenomenon, it is interesting to compare the
behavior of Li in CEMP-s, CEMP-rs, and CEMP-no stars.
The last column of Table \ref{tab:Abundances} reports the stellar
classification as CEMP-s, CEMP-rs,or CEMP-no, for the stars of our sample based
on the Ba and Eu abundances, as defined in \citet{Masseron2010}. The
classification for the stars mentioned in Table~\ref{tab:extended} can be found
in \citet{Masseron2010}. Fig.~\ref{fig:Livssubclasses} shows the distribution
of Li abundances for stars in each subclass. There are Li-normal and Li-depleted stars in
both the CEMP-s and CEMP-rs subclasses, which suggests either that the carbon
for each subclass comes from a similar source or that effects in the CEMP star,
such as extra mixing, dominate over the Li signature of the polluting material.
All of the CEMP-no stars in the extended sample are Li-depleted. The origin of
CEMP-no stars is still uncertain -- their peculiar abundances come either from a
non-$s$-process-element producing AGB mass transfer or from C-rich gas coming
from a previous generation of massive stars \citep[see, e.g., ][ for more
details]{Masseron2010}. If the C comes from pollution by the winds of rapidly
rotating massive stars, it is not clear if the Li abundances can discriminate
between the scenarios, because the yields given by \citet{Meynet2006} are
compatible with AGB yields down to the point that ``almost no distinction could
be made'' \citep{Meynet2006}. Fig.~\ref{fig:LivsC+N_dilu_models} shows that with
high dilution factors ($\sim$50\%), the wind models can explain some Li-depleted
CEMP-no stars with moderate carbon enhancements. The lack of Li-normal stars
among the CEMP-no stars may also be the result of the combination of their
metallicities and our definition of ``Li-normal''. \citet{Aoki2007} and
\citet{Masseron2010} noted that CEMP-no stars are on average more metal-poor
([Fe/H]$\lesssim -2.7$) than CEMP stars in other subclasses.
\citet{Bonifacio2007}, \cite{Aoki2009}, and \citet{Sbordone2010} found that 
non-C-rich metal-poor stars with [Fe/H] $< -3.0$ appear to exhibit a decrease in Li
abundance compared to the Spite plateau. Hence, the fact that we do not see
Li-normal stars among CEMP-no stars might also be affected by our conservative
definition of Li-normal stars, which does not encompass stars with
$\log_\epsilon(Li) < 2.0$.  However, in
the end, our sample of the CEMP-no stars is small; more Li data for this category could
reveal Li-normal stars here as well.  

The presence of Li-normal stars among CEMP-rs stars is informative for another
reason. As observed by \citet{Masseron2010}, some CEMP stars, mostly in the
CEMP-rs subclass, lie off the standard evolutionary tracks and are more luminous
or bluer than expected from isochrones of the appropriate metallicities. It is
not yet clear why these stars show such properties, but some of the scenarios
for forming field blue stragglers imply mass transfer of material, as is the
case for CEMP-rs stars. 
CEMP-rs stars show the highest enhancements of C and neutron-capture elements in
their atmospheres. This could drive efficient thermohaline mixing, which drags
the C-enriched material down to the core and boosts the luminosity of the stars
through the CNO cycle \citep{Stancliffe2007}. In any possible scenarios, Li
should have been heavily destroyed. However, the upper panel of
Fig.~\ref{fig:Livssubclasses} shows that there are Li-normal stars in the
CEMP-rs class. Therefore, the reason why some CEMP-rs stars are not located on
the regular track of the HR diagram remains unsolved.

\section{Conclusions} 

We have used Li abundances and carbon isotope ratios to test the origin of
carbon and the presence of mixing and dilution in CEMP stars. We observe a large
range of Li abundances in CEMP stars. If moderate amounts of AGB material are
transferred to the CEMP star, then dilution appears to explain Li-normal stars.
Although we cannot completely rule out that some Li may have been produced in
a former AGB companion, it is definitely not required to invoke that process to
explain the CEMP stars with Li abundances close to the Spite plateau value. In
contrast, it appears difficult to explain the CEMP stars where the Li is
depleted without additional Li depletion in the CEMP star itself. Hence, extra
mixing is required. 
Although thermohaline mixing is indeed an obvious candidate to
explain Li depletion in some CEMP stars, we did not find any correlation between
the CEMP properties extensively explored by \citet{Stancliffe2009Li} and Li
depletion. However, rotation seems to play a prominent role for some CEMP
stars. We predict that the fastest rotating stars (which
correspond to the stars with the largest broadening profiles), should exhibit the
largest Li depletion. This result is in line with the recent findings of
\citet{Smiljanic2010} and \citet{CantoMartins2011}, showing that the inclusion of rotation in the
evolutionary models can reproduce Li depletion in open clusters. 

Because the current sample of CEMP stars with Li measurements is
still small, we could not definitively infer a dependency of the rotational
velocity on the amount of Li depletion. Larger samples and more systematic radial
velocity monitoring is required to confirm these results. 

\acknowledgments

We thank T. Dermine for discussions on binary accretion and N. Behahra for
providing spectra. We acknowledge support from NSF AST-0607482 and AST-0707948.
TCB acknowledges partial support from grants PHY 02-16783 and PHY 08-22648:
Physics Frontiers Center/Joint Institute for Nuclear Astrophysics (JINA),
awarded by the U.S. National Science Foundation. NC acknowledges support
by Sonderforschungsbereich SFB 881 ``The Milky Way System'' (subproject A4)
of the German Research Foundation (DFG).

\bibliography{thomas_ApJ}

\end{document}